%% file: main.tex
\documentclass[sigconf, usenames, dvipsnames]{acmart}

\usepackage{dblfloatfix} % for figure placement at bottom
\usepackage{booktabs} % multi column
\usepackage{arydshln}
\usepackage{subfig}
\usepackage{multirow}
\usepackage{xcolor}

\usepackage{listings}
\usepackage{xcolor}
\usepackage{booktabs} % Include in preamble

\AtBeginDocument{%
  \providecommand\BibTeX{{%
    \normalfont B\kern-0.5em{\scshape i\kern-0.25em b}\kern-0.8em\TeX}}}

\newcommand{\system}{Guided Reality }

\begin{document}
\pagenumbering{arabic}
\title{Guided Reality: Generating Visually-Enriched AR Task Guidance with LLMs and Vision Models}
% Fully Automated Generation of AR Task Guidance with Rich Visual Augmentations
% Guided Reality: Generating Visual Task Guidance in Augmented Reality

% other options: Generating Visual Guidance in Augmented Reality with LLM?

\author{Ada Yi Zhao}
\orcid{0009-0009-5739-2176}
\affiliation{%
  \institution{University of Colorado Boulder}
  \streetaddress{Address}
  \city{Boulder}
  \country{USA}}
\email{ada.zhao@colorado.edu}

\author{Aditya Gunturu}
\orcid{0009-0000-5126-6579}
\affiliation{%
  \institution{University of Calgary}
  \streetaddress{Address}
  \city{Calgary}
  \country{Canada}}
\affiliation{%
  \institution{University of Colorado Boulder}
  \streetaddress{Address}
  \city{Boulder}
  \country{USA}}  
\email{aditya.gunturu@ucalgary.ca}

\author{Ellen Yi-Luen Do}
\orcid{0000-0002-9948-6375}
\affiliation{%
  \institution{University of Colorado Boulder}
  \streetaddress{Address}
  \city{Boulder}
  \country{USA}}
\email{ellen.do@colorado.edu}

\author{Ryo Suzuki}
\orcid{0000-0003-3294-9555}
\affiliation{%
  \institution{University of Colorado Boulder}
  \streetaddress{Address}
  \city{Boulder}
  \country{USA}}
\email{ryo.suzuki@colorado.edu}

\renewcommand{\shortauthors}{Zhao, et al.}

\copyrightyear{2025}
\acmYear{2025}
\setcopyright{acmlicensed}\acmConference[UIST '25]{The 38th Annual ACM Symposium on User Interface Software and Technology}{September 28-October 1, 2025}{Busan, Republic of Korea}
\acmBooktitle{The 38th Annual ACM Symposium on User Interface Software and Technology (UIST '25), September 28-October 1, 2025, Busan, Republic of Korea}
\acmDOI{10.1145/3746059.3747784}
\acmISBN{979-8-4007-2037-6/2025/09}

\input{0-abstract}

\begin{teaserfigure}
\centering
\includegraphics[width=0.24\linewidth]{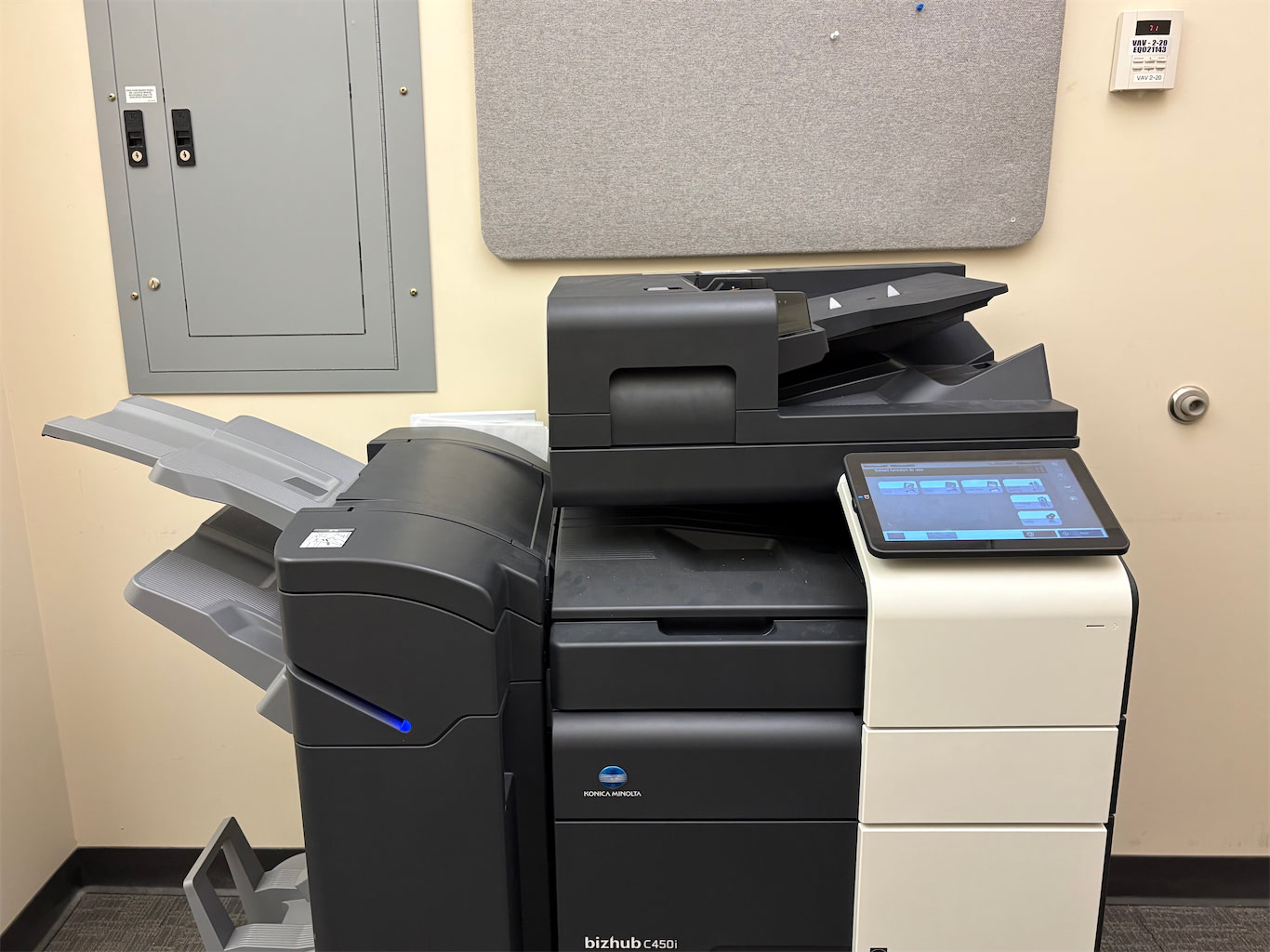}
\includegraphics[width=0.24\linewidth]{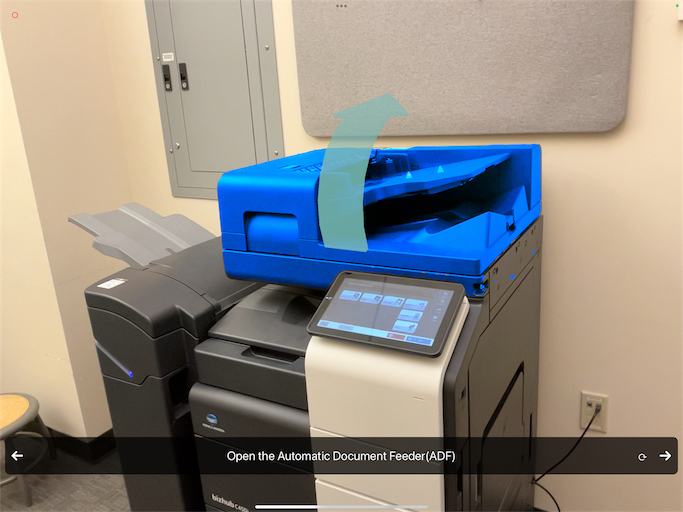}
\includegraphics[width=0.24\linewidth]{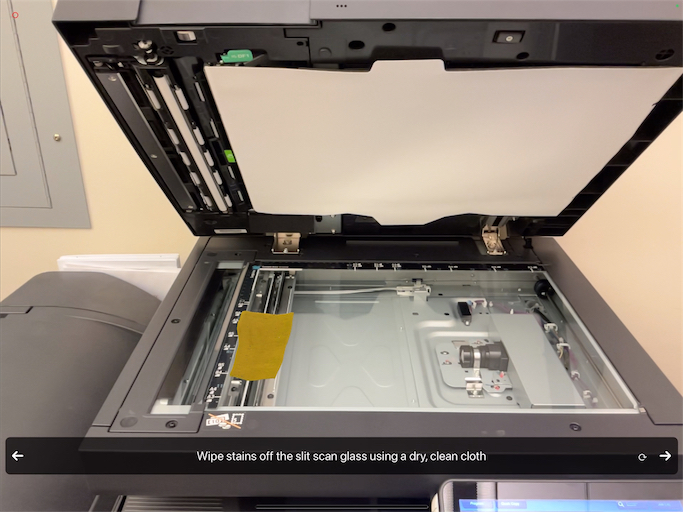}
\includegraphics[width=0.24\linewidth]{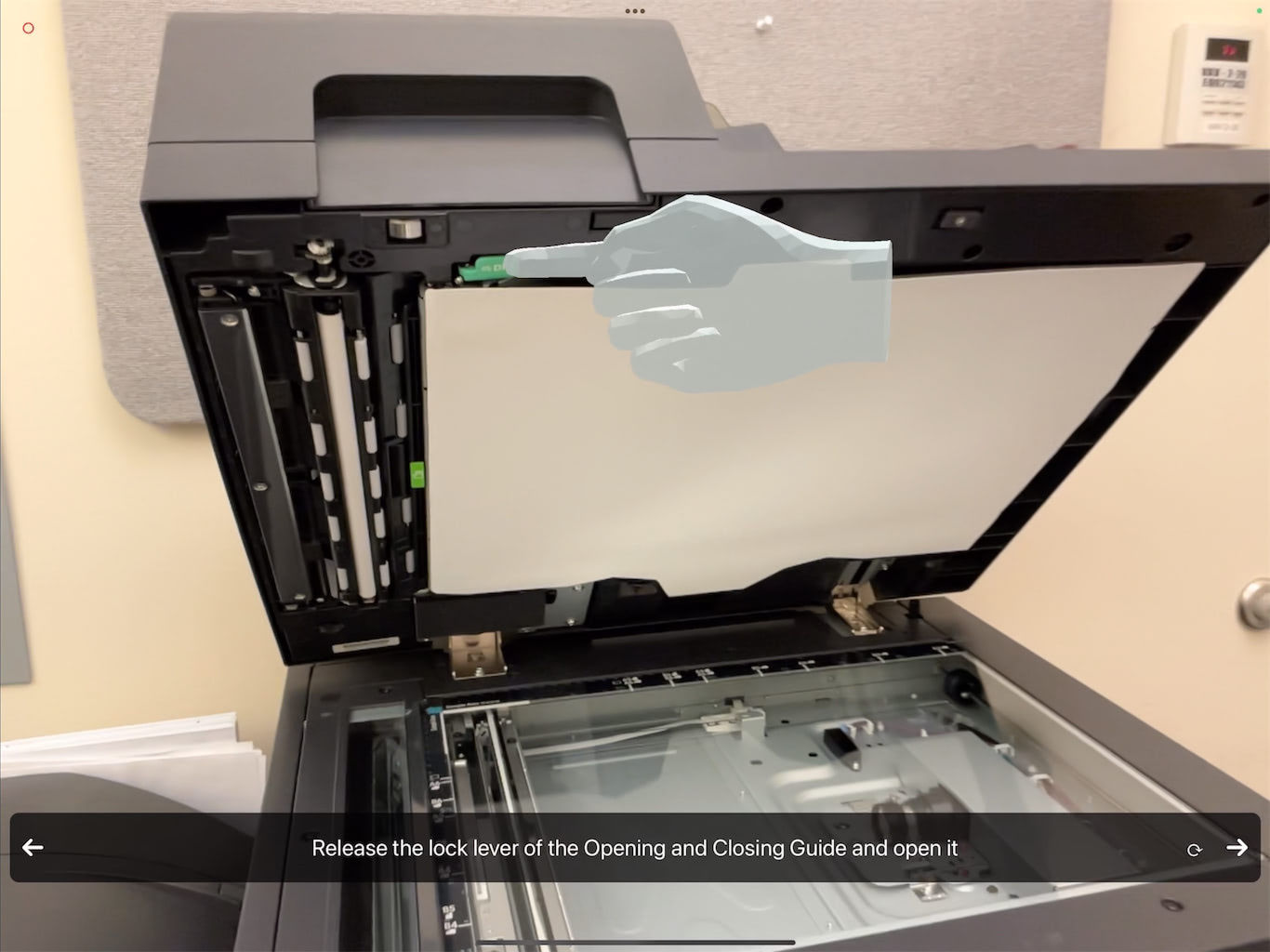}
\\[0.05cm]
\includegraphics[width=0.24\linewidth]{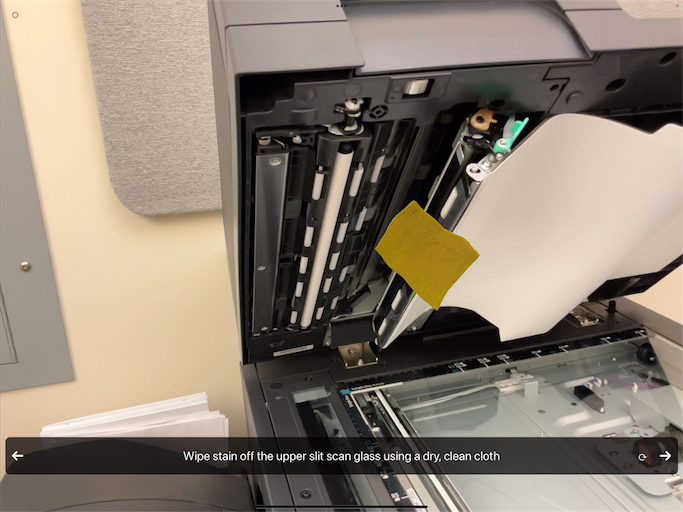}
\includegraphics[width=0.24\linewidth]{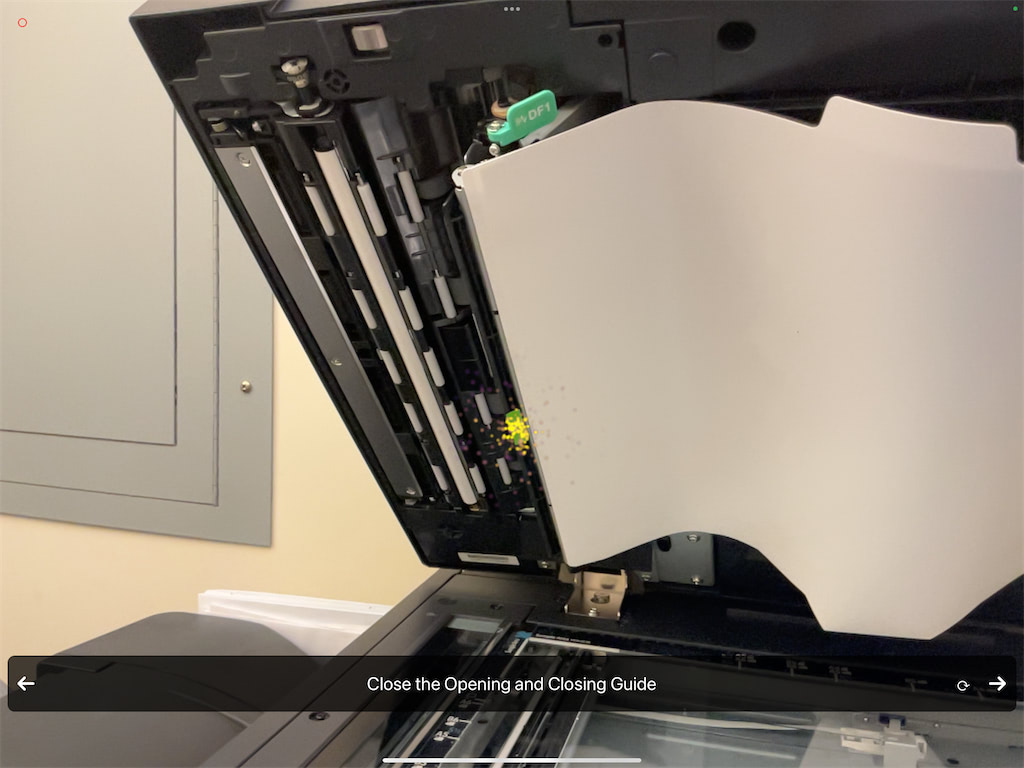}
\includegraphics[width=0.24\linewidth]{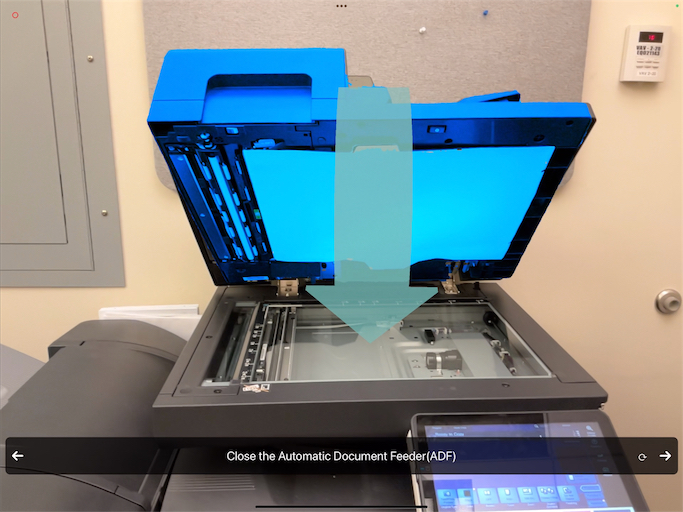}
\includegraphics[width=0.24\linewidth]{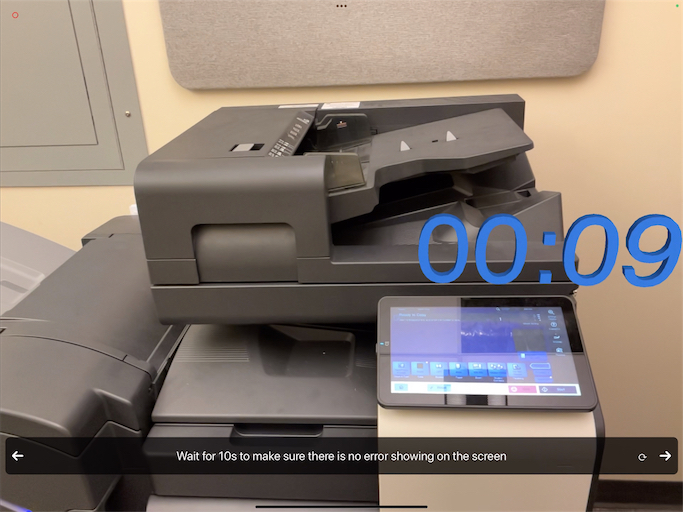}
\caption{The Guided Reality system instructs users to clean the scanning area of an office printer. The image demonstrates (1) setup, (2) opening the Automatic Document Feeder (ADF) with \textit{movement indication of segmented object}, (3) wiping the lower glass with an \textit{animated tool}, (4) releasing the lever with a \textit{hand gesture}, (5) cleaning the upper glass with \textit{animated tool}, (6) closing the guide with \textit{particle highlight}, (7) closing the ADF with \textit{movement indication of segmented object}, and (8) waiting 10 seconds to check for errors with a \textit{contextual widget of a timer}.}
\label{fig:teaser}
\end{teaserfigure}

% \includegraphics[width=\textwidth]{figures/Frame 2.png}
% \caption{The eight types of visual indications that we proposed to help users understand the text instruction in Augmented Reality}

\maketitle

\input{1-introduction}
\input{2-related-work}
\input{3-formative-study}
\input{4-system-design}

\input{5-implementation}

\input{6-user-study}

\input{7-future-work}

\input{8-conclusion}

\input{acknowledgements}

\balance
\bibliographystyle{ACM-Reference-Format}
\bibliography{references}

\appendix

\section{Appendix}

\subsection{Structure Task Plan Prompt}
\label{appendix:taskPlanPromp}

You will be given an image showing the user’s current view and a question about completing a physical task. Act as an assistant that outputs a list of step-by-step instructions for the user to follow. For each step, you will also select the most appropriate visualization type (from five provided options) and include any related visual details. The output must be in JSON format.

The top-level JSON field should be \texttt{"instructions"}, whose value is a list of steps. Each step is a JSON object with the following fields:

\begin{itemize}
    \item \texttt{"instruction"}: A short, precise text description of the step.
    \item \texttt{"key\_components"}: A list identifying the relevant components. The first item must be the object the user will interact with, described as: \emph{component property + component name}. Properties can include relative position, color, or visible text to help distinguish it from similar components in the image. More detail may be required depending on the chosen visual type.
    \item \texttt{"visual\_type"}: An integer indicating the best-fitting visualization type out of 5 visual types for this step. First check if the instruction is just waiting. If so, select type 5. Otherwise, if it uses an external tool, use type 4. If done by bare hand, check for a specific gesture to select type 3, then check for movement to select type 2. If none apply, default to type 1. 
\end{itemize}

The five visual types are described below:

\noindent Type 1: Highlighting the key area

The objective is to help users locate the key component. A sample use case is pressing a button. Example JSON:

\begin{verbatim}
{
  "instruction": "press start button on the rice
                 cooker",
  "visual_type": 1,
  "key_components": [
    "The orange Start button"
  ]
}
\end{verbatim}

\noindent Type 2: Indicating key component movement  

The objective is to show the component’s movement using start and end points. A sample use case is returning the basket to the air fryer or moving a chess piece. The first item in \texttt{key\_components} is the component to move; the second specifies the movement type: either translation or rotation. Example JSON:
\begin{verbatim}
{
  "instruction": "Return the basket to the air fryer
                  to resume cooking",
  "visual_type": 2,
  "key_components": [
    "Air fryer basket", "translation"
  ]
}
\end{verbatim}

\noindent Type 3: Hand gesture  

The objective is to show the specific hand gestures required to complete a step. The information needed includes the \texttt{key\_component} to be manipulated and the type of gesture used to operate the machine. Choose from the valid gestures include: poke, hook, palm press, grip, cylindrical grasp, and pinch. Example JSON:
\begin{verbatim}
{
  "instruction": "Pull the filament out",
  "visual_type": 3,
  "key_components": [
    "filament on top of nozzle", "pinch"
  ]
}
\end{verbatim}

\noindent Type 4: External tool with movement indication  
The objective is to show what tool is required and the movement the user should perform with it. Sample use cases include tightening a screw with a screwdriver or mixing ingredients with a whisk. The \texttt{key\_components} field includes three items: the component or area to interact with, the tool's movement (choose from "up and down", "left and right", "rotate", "clockwise", or "counterclockwise"), and the name of the tool. Example JSON:
\begin{verbatim}
{
  "instruction": "Mix the ingredients with a whisk",
  "visual_type": 4,
  "key_components": [
    "Mixing bowl", "rotate", "whisk"
  ]
}
\end{verbatim}

\noindent Type 5: Show a dynamic widget  

A use case is when the user needs to wait for a certain amount of time. The \texttt{key\_components} field includes two items: the area or component to wait for, and the wait time in \texttt{mm:ss} format. Example JSON:
\begin{verbatim}
{
  "instruction": "Let the food stand for 30s",
  "visual_type": 5,
  "key_components": [
    "Mixing bowl", "00:30"
  ]
}
\end{verbatim}

\subsection{2D Bounding Box Prompt}
\label{appendix:boxPromp}

I want to identify where the \texttt{key\_component} is in the image. Output both the object name and position using this JSON format: 

\texttt{\{\{name: key\_component\_name, pos: [y\_min, x\_min, y\_max, x\_max]\}\}}.

\subsection{Translational Movement Prompt}
\label{appendix:translationalPromp}

I want to identify where the \texttt{key\_component} is in the image. Please also identify where this \texttt{key\_component} will end up after the instruction \texttt{instruction}. Output the object name, original position and end target position using this JSON format: 

\texttt{\{\{name: key\_component\_name, pos: [y\_min, x\_min, y\_max, x\_max], target\_pos: [xEnd, yEnd]\}\}}.

\subsection{Rotation Information Prompt}
\label{appendix:rotationalPrompt}

Given an instruction \texttt{instruction}, please identify how the \texttt{key\_component} rotates in the second image. Refer to the first image to set up the spatial axis, with $x$ pointing rightward in the photo, $y$ pointing physically upward, and $z$ pointing toward you. Identify the rotational axis ($x$, $y$, or $z$). Always look from the positive side and determine the rotation direction (clockwise, CW, or counterclockwise, CCW). Output the information using this JSON format:

\texttt{\{\{rotation: [axis, direction]\}\}}

\end{document}

%% file: 0-abstract.tex
\begin{abstract}
Large language models (LLMs) have enabled the automatic generation of step-by-step augmented reality (AR) instructions for a wide range of physical tasks. However, existing LLM-based AR guidance often lacks \textit{rich visual augmentations} to effectively embed instructions into spatial context for a better user understanding. We present Guided Reality, a fully automated AR system that generates embedded and dynamic visual guidance based on step-by-step instructions. Our system integrates LLMs and vision models to: 1) generate multi-step instructions from user queries, 2) identify appropriate types of visual guidance, 3) extract spatial information about key interaction points in the real world, and 4) embed visual guidance in physical space to support task execution. Drawing from a corpus of user manuals, we define five categories of visual guidance and propose an identification strategy based on the current step. We evaluate the system through a user study (N=16), completing real-world tasks and exploring the system in the wild. Additionally, four instructors shared insights on how Guided Reality could be integrated into their training workflows.
\end{abstract}

\begin{CCSXML}
<ccs2012>
   <concept>
       <concept_id>10003120.10003121.10003124.10010392</concept_id>
       <concept_desc>Human-centered computing~Mixed / augmented reality</concept_desc>
       <concept_significance>500</concept_significance>
   </concept>
 </ccs2012>
\end{CCSXML}

\ccsdesc[500]{Human-centered computing~Mixed / augmented reality}
% http://dl.acm.org/ccs.cfm

\keywords{Augmented Reality; Mixed Reality; LLMs; Task Guidance}

%% file: 1-introduction.tex
\section{Introduction}
Augmented reality (AR) has become an effective medium for delivering step-by-step instructions in various domains, from object assembly~\cite{funk2016interactive, yamaguchi2020video} to machine operations~\cite{cao2020exploratory}. By overlaying information directly in the user’s field of view, AR guidance can significantly reduce cognitive load~\cite{obermair2020maintenance}, lower error rates~\cite{pietschmann2023quantifying}, and provide a more engaging learning experience~\cite{zhang2025following}. Recent advances in large language models (LLMs) further enhance AR instruction by \textit{automatically generating on-demand guidance} that adapts to individual tasks and contexts~\cite{bohus2024sigma, shi2025caring}. This approach shows great promise as it removes the need for manually authored or pre-defined instruction content, offering significant flexibility and scalability for diverse real-world use.

Despite these advancements, most current LLM-based AR guidance systems~\cite{stover2024taggar, srinidhi2024xair, shi2025caring} lack the ability to generate \textit{rich visual augmentations}---a variety of spatially embedded graphical cues such as animated 3D overlays, diverse hand gestures, tool interactions, and highlighted motion paths. Prior research shows that these rich visual augmentations could significantly improve users’ understanding and task performance~\cite{cao2020exploratory, huang2021adaptutar, chidambaram2021processar}, but existing LLM-driven AR guidance often limited to the generation of basic spatial highlights~\cite{bohus2024sigma, srinidhi2024xair} or limited motion guidance~\cite{shi2025caring}.

In this paper, we present Guided Reality, a fully automated AR system that generates and embeds visually rich, spatially grounded task guidance. For example, consider a user cleaning the scanning area of an office printer (Figure~\ref{fig:teaser}): our system translates step-by-step textual instructions into animated and contextual visual guidance---1) a motion cue for opening the automatic document feeder, 2) an animated cloth cleaning the surface, 3) a hook hand gesture to release the lever, 4) a highlighted area to push to close the opening and closing guide, and 5) a contextual widget to countdown the time. As shown in Figure~\ref{fig:teaser}, Guided Reality enables to generate and embed expressive and situated visual instructions.

To achieve this level of guidance, the system must determine the appropriate type of visual guidance for each instruction step, generate visual content tailored to the physical task, and embed this content within the user’s environment in real-time. These challenges map to two key research goals: 1) \textbf{\textit{Defining a visual guidance strategy}} that identifies what kind of visuals should accompany each instruction step; and 2) \textbf{\textit{Automatically generating and placing those visuals}} in AR, without manual intervention.

To address the first goal, we conducted an analysis of existing user manuals to study how visual elements are used to support task instructions. From this, we identified five common types of visual guidance: 1) highlighting a key component, 2) indicating movement, 3) demonstrating a hand gesture, 4) showing a 3D tool with motion, and 5) displaying a contextual widget like a timer. We use this framework to guide system design and to cover a wide range of applications domains, including household appliance operation, lab equipment, and recreational activities.

To address the second goal, we developed a fully automated AR pipeline. We use ChatGPT to interpret user queries and generate a structured task plan as JSON, including textual instructions, identified visual guidance types, and descriptions of key components. Based on the visual type of each step, we extract a curated collection of 3D tools or produce on-the-fly visual content using Segment Anything~\cite{ravi2024sam}. Finally, we localize and render these visuals in space by combining 2D bounding box of the key component from Google Gemini with spatial data from the device. This allows us to place the guidance accurately within the 3D scene and anchor it to real-world objects, enabling seamless task execution.

We evaluated our system through two user studies. The first study involved 16 participants and consisted of two parts: a controlled study assessing the effectiveness of our visual guidance strategy, and an in-the-wild session to examine how the system performs in real-world scenarios. The second study consisted of expert interviews with four instructors, which provided deeper insights into the benefits and limitations of our approach from a teaching and training perspective. Together, the results indicate that Guided Reality enhances user learning experiences and reveal future directions for generated AR task guidance.

In summary, our contributions include:
\begin{enumerate}
\item Guided Reality, a fully automated AR system that generates and embeds visually rich, spatially grounded task guidance.
\item A visual guidance analysis and strategy that identifies appropriate visual guidance type given a text instruction.
\item Insights from the two user evaluations from both learners (N=16) and instructors (N=4).
\end{enumerate}

%% file: 2-related-work.tex
\section{Related Work}

\subsection{AR Guidance and Visual Representations}

AR-based task guidance has been extensively explored across a variety of domains, including machine operations~\cite{cao2020exploratory}, object assembly~\cite{funk2016interactive, yamaguchi2020video}, and industrial maintenance~\cite{obermair2020maintenance}. Rather than simply projecting textual instructions from paper manuals into physical space, AR systems can overlay rich visual guidance directly within the user's workspace~\cite{liu2023instrumentar}. For example, systems like AdapTutAR~\cite{huang2021adaptutar} demonstrate how \emph{rich visual affordances}—such as avatars, arrows, and animated highlights—can enhance comprehension and reduce errors. A wide range of such affordances has been proposed, including object highlighting~\cite{speicher2019exploring, cao2022mobiletutar}, virtual hands~\cite{haltner2023comparative, faridan2023chameleoncontrol, chidambaram2021processar, bai2020user, kim2019evaluating}, animated avatars~\cite{huang2021adaptutar, chidambaram2022editar, cao2020exploratory}, 3D model overlays~\cite{chidambaram2021processar}, spatially anchored text annotations~\cite{huang2021adaptutar, zhang2025following}, and point-cloud replays~\cite{cao2022mobiletutar, thoravi2019loki}. Additional visual elements such as arrows~\cite{wu2025rubikon, stover2024taggar}, overlaid annotations~\cite{mohr2017retargeting}, menus~\cite{dogan2024augmented}, and navigation cues~\cite{yuan2024augmented, cao2022mobiletutar, kim2019evaluating} have also been employed to support interactive, step-based tasks. This rich variety of visual augmentations has demonstrated numerous benefits, including reductions in task time~\cite{polvi2017handheld, pietschmann2023quantifying}, error rates~\cite{pietschmann2023quantifying}, and cognitive load~\cite{obermair2020maintenance}, as well as increased user satisfaction~\cite{cao2020exploratory}. Pietschmann et al.~\cite{pietschmann2022extended} categorized such visual guidance into three types: 1) gaze guidance to navigate where to look, 2) object identification to indicate what to focus on, and 3) action guidance to assist how to act. While combining all types of visual guidance may seem beneficial, Cao et al.~\cite{cao2020exploratory} cautioned that doing so is not always effective. Huang et al.~\cite{huang2021adaptutar} further demonstrated that visual guidance can be adapted to users' expertise levels in real time. Yet, no prior research has examined which types of visual guidance are most suitable for different kinds of instructional content.

\subsection{Authoring Tools for AR Instructions}
While AR guidance demonstrates effectiveness, most existing systems rely on pre-defined instructions or require manual configuration by experts to specify both the visual form and spatial placement of instructions~\cite{MicrosoftDynamics}. This motivates the need for authoring tools that simplify the creation of AR instructions. For instance, systems like ProcessAR~\cite{chidambaram2021processar}, CAPturAR~\cite{wang2020capturar}, TutorialLens~\cite{kong2021tutoriallens}, and InstruMentAR~\cite{liu2023instrumentar} explore authoring workflows that allow users to create AR guidance through task demonstrations and multimodal recordings. Beyond live demonstrations, several systems leverage recorded demonstrations as a basis for AR tutorial generation. For example, MobileTutAR~\cite{cao2022mobiletutar} and EditAR~\cite{chidambaram2022editar} extract expert task performances from 2D video segmentation or 3D digital twins to generate replayable AR instructions. Other systems, such as eXplainMR~\cite{wang2025explainmr}, GesturAR~\cite{wang2021gesturar}, and ScalAR~\cite{qian2022scalar}, explore complementary approaches including automatic explanation generation~\cite{wang2025explainmr}, gesture-driven programming~\cite{wang2021gesturar}, and semantic adaptation~\cite{qian2022scalar}. While these approaches mark significant progress, they still often require manual selection and placement of visual guidance types.

\subsection{AI for AR Task Guidance}
Recent work explores the use of machine learning and AI to automate this process~\cite{stover2024taggar, srinidhi2024xair, keelawat2024transforming}. For instance, PaperToPlace~\cite{chen2023papertoplace} optimizes the placement of text instruction from printed manual to mixed-reality experiences automatically without occlusion. Most relevant to our work, TAGGAR~\cite{stover2024taggar}, CARING-AI~\cite{shi2025caring}, and XaiR~\cite{srinidhi2024xair} leverage multimodal LLM and generative AI pipelines for AR task guidance. TAGGAR~\cite{stover2024taggar} uses vision-language models to extract actions and object references from natural language and images, placing predefined hand animations at localized regions. CARING-AI~\cite{shi2025caring} enables authors to generate humanoid avatar animations from structured textual instructions using generative motion models and contextual scanning. XaiR~\cite{srinidhi2024xair} integrates LLMs with egocentric video and 3D spatial data to provide progress-aware task guidance through anchored AR overlays during live instruction.

Our work pushes the boundaries of these LLM-based AR guidance systems in two ways. First, we expand the visual guidance types based on a taxonomy analysis of user manuals, beyond basic spatial highlights~\cite{bohus2024sigma, srinidhi2024xair} or limited motion guidance~\cite{shi2025caring}. Second, we explore a fully-automated pipeline to identify, generate, and embed such rich visual guidance, eliminating the need for predefined operator-authored content \cite{stover2024taggar}. These visual forms are dynamically grounded in the user's live environment using real-time computer vision models like Segment Anything~\cite{ravi2024sam} and depth estimation. With these, we aim to advance and demonstrate visually-rich, fully automated LLM-based AR guidance across diverse physical tasks without requiring explicit demonstrations and pre-authored assets.

\subsection{LLMs for AR Interfaces}
Beyond task guidance, researchers have also explored incorporating LLMs and generative AI into general AR interfaces. For example, XR-Objects~\cite{dogan2024augmented} introduces object-driven contextual menus powered by multi-modal LLMs, allowing analog objects to serve as interactive digital interfaces. RealitySummary~\cite{gunturu2024realitysummary} leverages LLMs to summarize users’ physical surroundings and generate high-level descriptions for context-aware interactions. GazePointAR~\cite{lee2024gazepointar} demonstrates how gaze and speech inputs can be combined with LLMs to produce just-in-time verbal guidance anchored to the user’s focus. While these systems highlight the rising potential of LLMs and generative AI in facilitating conversational, adaptive, and contextually responsive AR experiences, most focus on information delivery or object-centric interfaces rather than procedural guidance. In contrast, \emph{Guided Reality} integrates instruction generation, visual selection, and real-time grounding into a unified pipeline, enabling the seamless creation and delivery of situated task guidance.

%% file: 3-formative-study.tex
\section{Strategy of Visual Guidance in AR}
\begin{figure*}
  \centering
  \includegraphics[width=2\columnwidth]{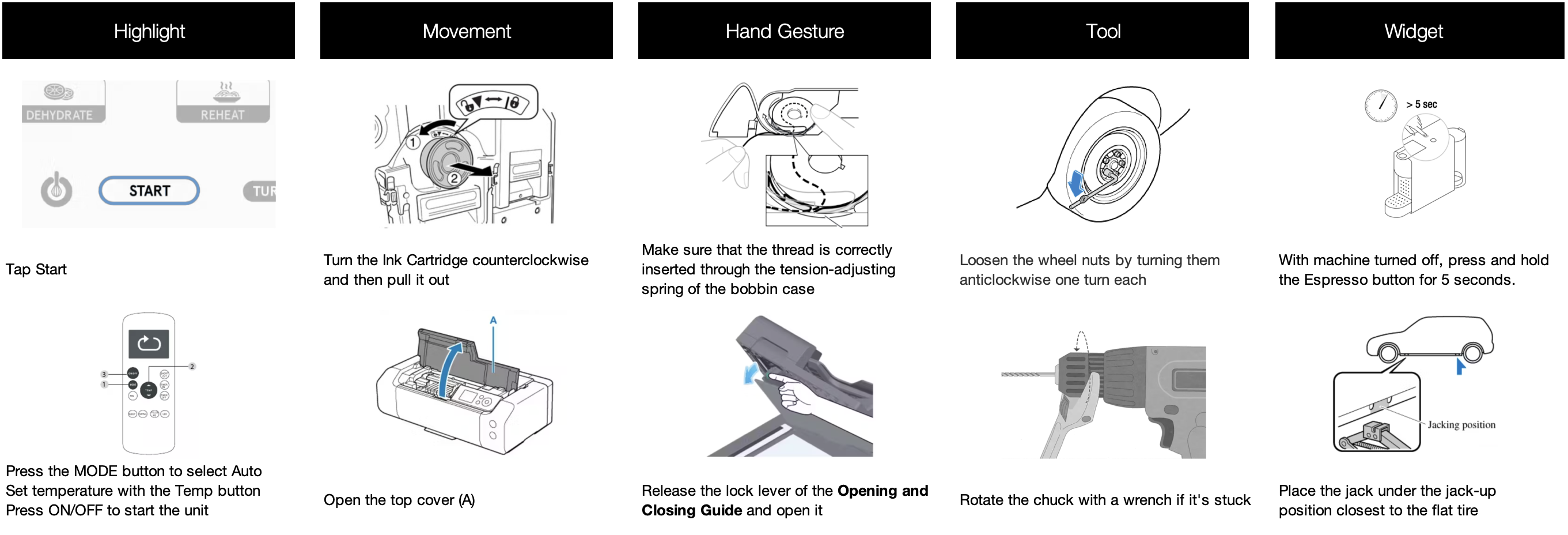}
    \caption{The design space of our system with examples from the user manuals we collected. Each step includes the text instruction and the paired image. From top to bottom and then left to right:~\cite{Gourmia} \textcopyright~Gourmia, ~\cite{controller} \textcopyright~TOSHIBA, ~\cite{cartridge} \textcopyright~RISO, ~\cite{Printer} \textcopyright~Canon, ~\cite{sewing} \textcopyright~brother, ~\cite{BigPrinter} \textcopyright~Konica Minolta, ~\cite{mazda} \textcopyright~Mazda, ~\cite{wrench} \textcopyright~wikiHow Jacob Pischer, ~\cite{coffee} \textcopyright~Nespresso, ~\cite{mazda} \textcopyright~Mazda,}
    \label{fig:design}
    \Description{A diagram showing}
\end{figure*}

To understand which visual elements are helpful for guiding task execution, we analyzed the relationship between text instructions and accompanying images in a diverse set of user manuals. This analysis revealed common visual strategies that support user comprehension and action. 

\subsection{Corpus and Methodology}
\label{sec:task_domains}
To inform the design of the \system system, we collected and analyzed user manuals that present step-by-step instructions through paired text and images. Our manual selection process is as follows. First, we identify the five task domains to include the diverse set of operations in everyday life, including 1) everyday household appliances, 2) workshop tools, 3) lab equipment, 4) safety and emergency devices, and 5) entertainment and consumer electronics. Second, given these domains, we identified common equipment and instructions for each category to collect the total of 16 instruction manuals for the diverse set. Our collected instruction includes handheld drill, sewing machine, air fryer, rice cooker, washer, 3D printer, laser cutter, and multimeter. Manuals consisting solely of visual instructions were excluded due to the lack of textual descriptions. Given these curated manuals spanning a broad range of domains, we identified 135 steps. After filtering out entries without clear image-text pairs, we retained 126 steps for our analysis.

Using this dataset, one of the authors conducted open coding of the paired instruction for each step to develop an initial set of categories, identifying preliminary dimensions of visual guidance. The entire team of authors reviewed and refined the proposed design space to ensure consistency and completeness. Two authors independently performed systematic coding on the entire dataset, followed by a collaborative reflection to resolve discrepancies and finalize the coding scheme.

\subsection{Design Space}
As illustrated in Figure~\ref{fig:design}, we defined the design space with \textbf{five common types of visual guidance} found in instructional materials: \textit{1) highlighting a component}, \textit{2) indicating movement}, \textit{3) demonstrating a hand gesture}, \textit{4) showing how to use a tool}, and \textit{5) displaying a contextual widget and information}. In the following sections, we describe each type of visual guidance in detail.

\subsubsection*{\textbf{D1. Highlighting a Component}}
User manuals often use visual cues such as bounding boxes to highlight the target area for simple, goal-oriented actions. These visuals serve primarily to help users quickly locate the relevant part of the object or interface. Such actions are typically intuitive and do not require detailed instructions on how to perform them, as the focus is on identifying what to interact with rather than how to do so. For example, instructions like \textit{``press the start button''} or \textit{``turn on the switch''} are often paired with bounding boxes around the corresponding component to draw the user's attention and facilitate quick execution.

\subsubsection*{\textbf{D2. Indicating Movement}}
When a step involves the physical movement of a component, user manuals typically highlight the component using a distinct visual cue, such as a colored overlay, to distinguish it from the surrounding elements and draw immediate visual attention. To indicate the direction of movement, arrows are often included, either linear or curved, depending on the motion. Additionally, when the task requires repositioning the component to a specific location, the intended end position is often illustrated to clarify the result of the action. For example, a printer manual may show the top cover highlighted with an arrow indicating an upward rotation to communicate the action \textit{``Open the top cover''}. Visual cues clarify both the component to be moved and the intended direction, helping users execute the step accurately.

\subsubsection*{\textbf{D3. Demonstrating a Hand Gesture}}
When an action involves a required or recommended hand gesture, user manuals typically include illustrations of the hand interacting with the object. These visuals communicate how the hand should be positioned and moved to perform the task accurately. Depicting hand gestures is especially important for actions involving fine motor control or coordinated use of both hands, where text alone may not sufficiently convey the necessary precision. For example, fire extinguisher instructions often show a hand squeezing the handle to activate the device. In more intricate tasks, such as threading a bobbin in a sewing machine, illustrations guide users by showing precise finger placement. These examples demonstrate how visualizing hand gestures enhances clarity in executing manual actions.
    
\subsubsection*{\textbf{D4. Showing How to Use a Tool}}
When an instruction involves the use of an external tool, user manuals typically include a visual representation of the tool to help users identify it. The tool is shown in the correct position and orientation relative to the target object to convey how it should be applied. To illustrate the required manipulation, directional arrows are commonly used to indicate motion. Typical tool operations include clockwise and counterclockwise rotation, rotational spinning, and linear movements such as up and down or left and right. For example, the instruction \textit{``Rotate the chuck with a wrench if it's stuck''} is accompanied by an image showing the wrench positioned on the chuck with a curved arrow indicating the direction of rotation. Similarly, the instruction “\textit{Wipe stains off the Slit Scan Glass using a dry, clean cloth}” is illustrated with a hand holding a cloth and an arrow showing the wiping motion across the surface. These visualizations help users understand both the required tool and the physical motion needed to perform the task accurately.

\subsubsection*{\textbf{D5. Displaying a Contextual Widget and Information}}
Certain tasks require waiting for a specific condition or a set duration before proceeding to the next step. For example, when baking, users might wait for an oven to preheat or for a specific cooking duration. Manuals often highlight indicators for such tasks. Although user manuals usually do not usually have an accompanied image for the wait instruction, previous research suggests that displaying a timer anchored at the object can assist users in tracking the time and managing time effectively \cite{dogan2024augmented, barquero2024understanding}.

%% file: 4-system-design.tex
\begin{figure*}[h!]
\centering
\includegraphics[width=0.24\linewidth]{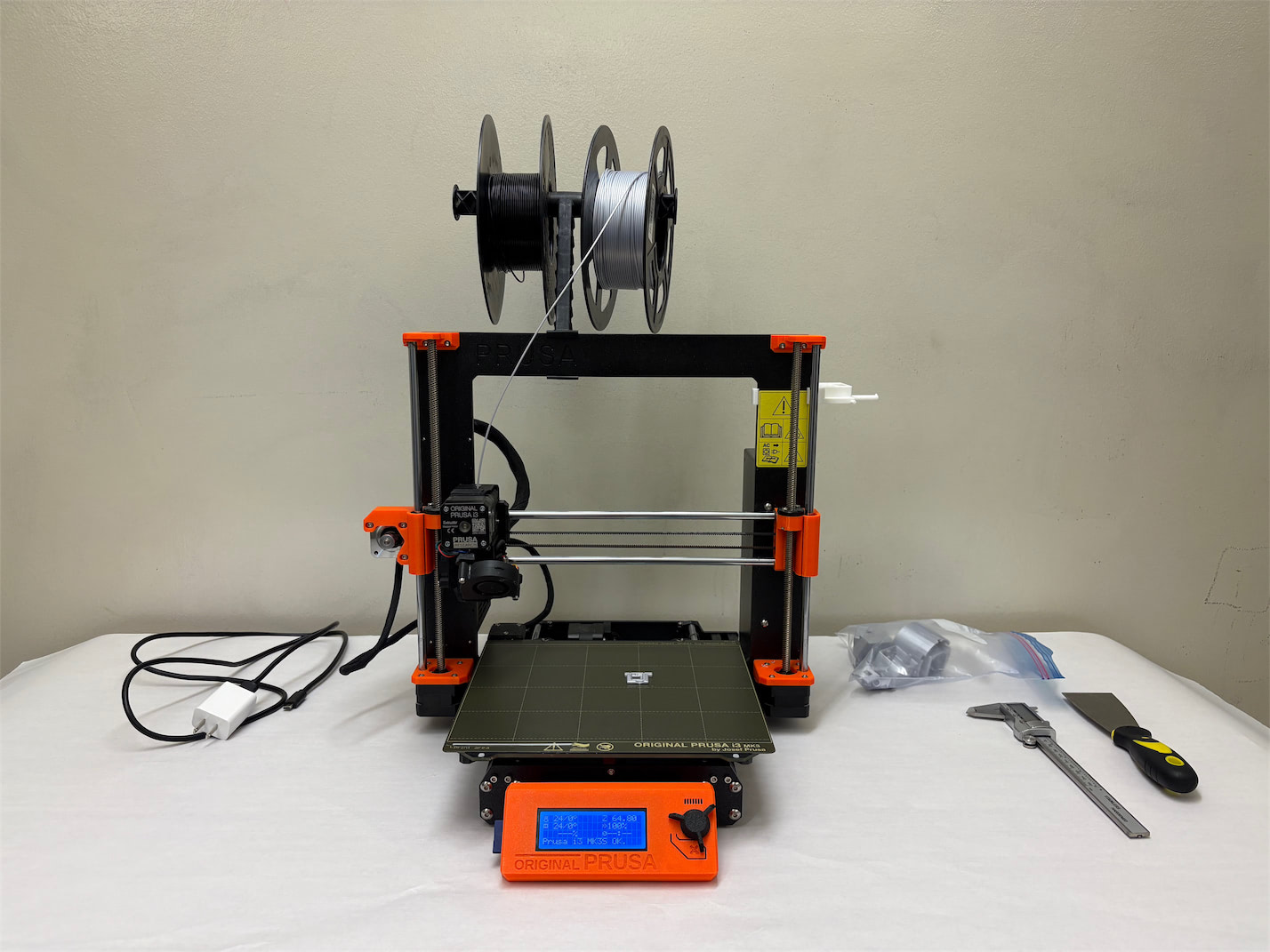}
\includegraphics[width=0.24\linewidth]{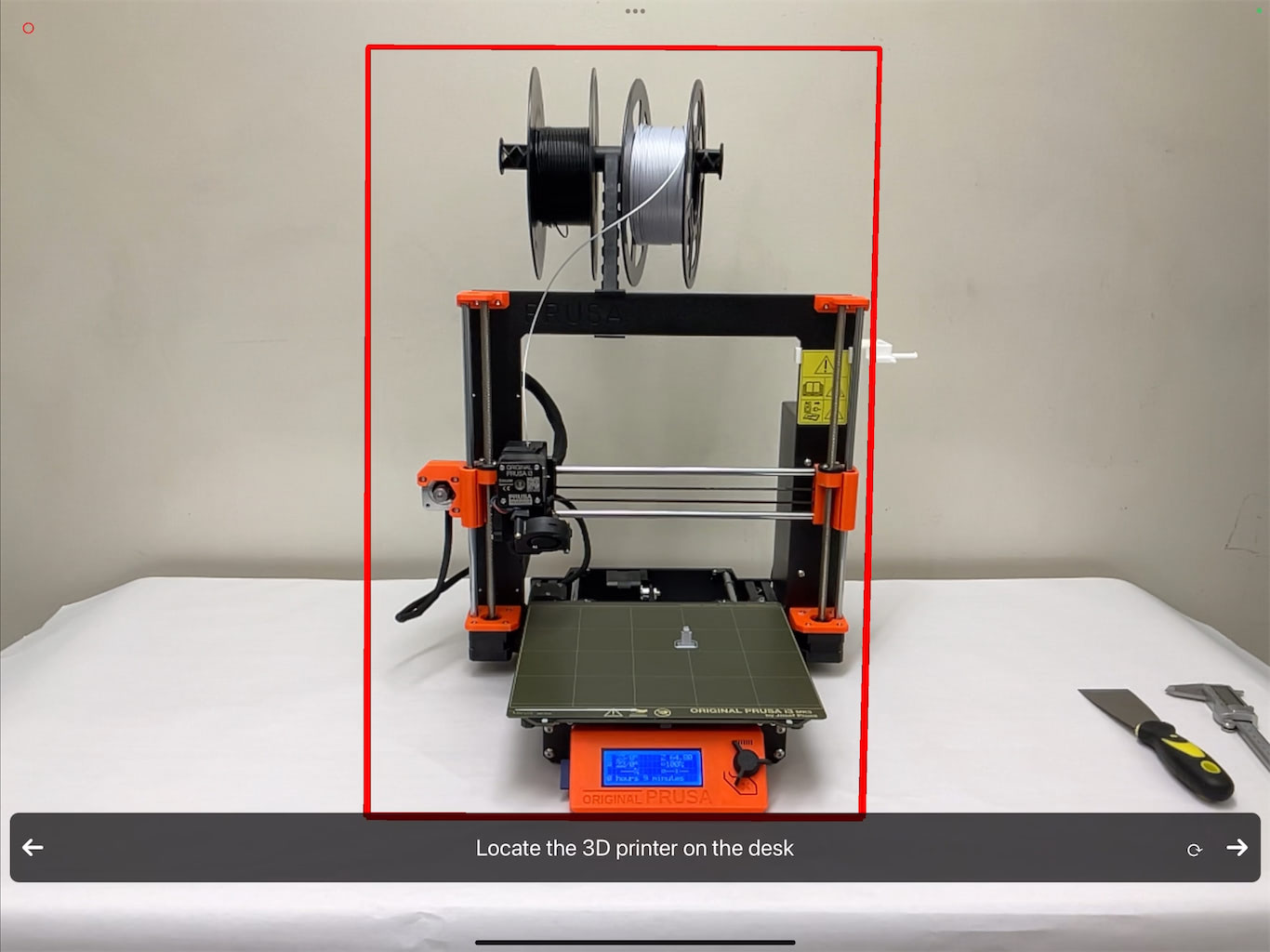}
\includegraphics[width=0.24\linewidth]{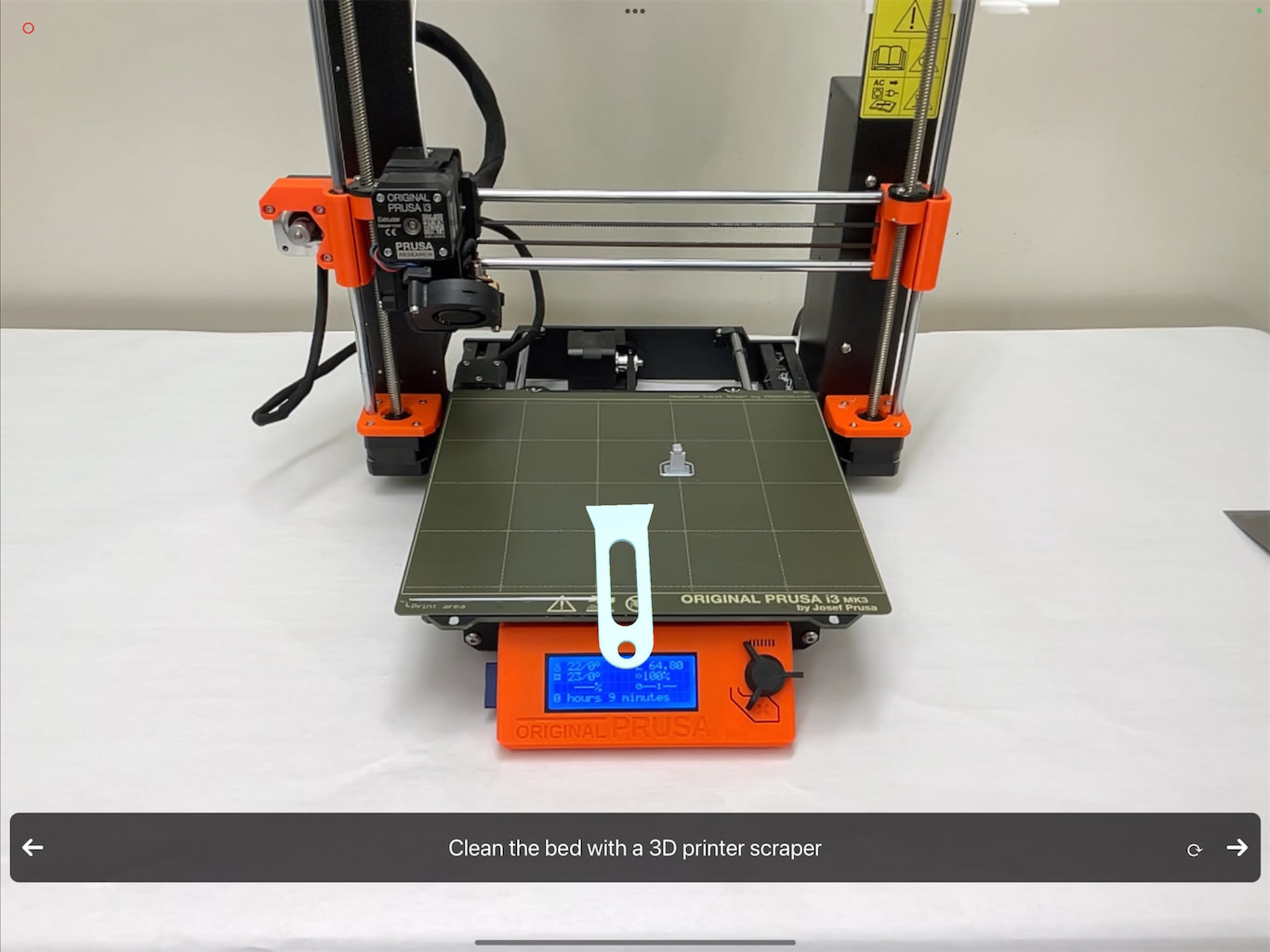}
\includegraphics[width=0.24\linewidth]{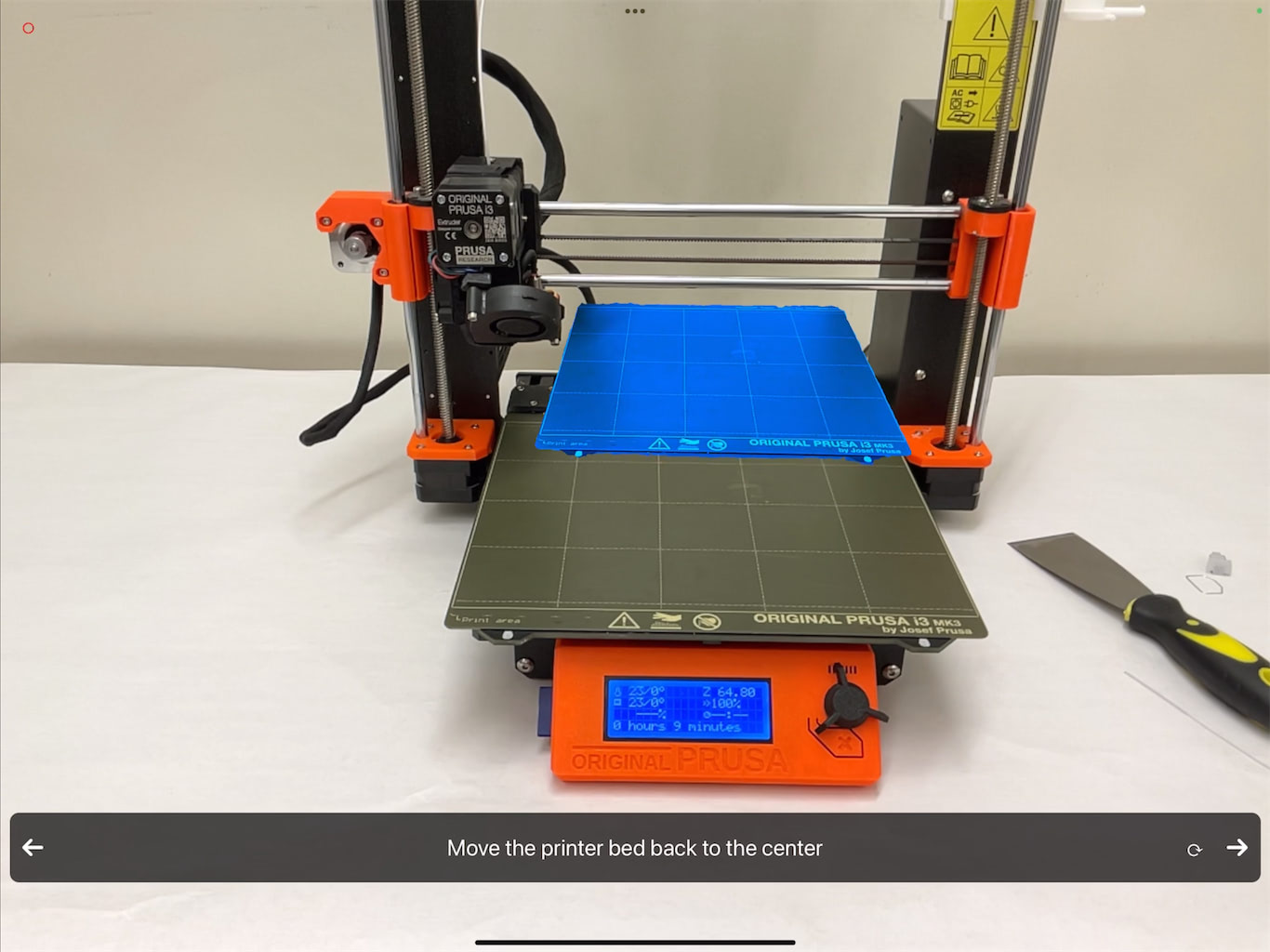}
\\[0.05cm]
\includegraphics[width=0.24\linewidth]{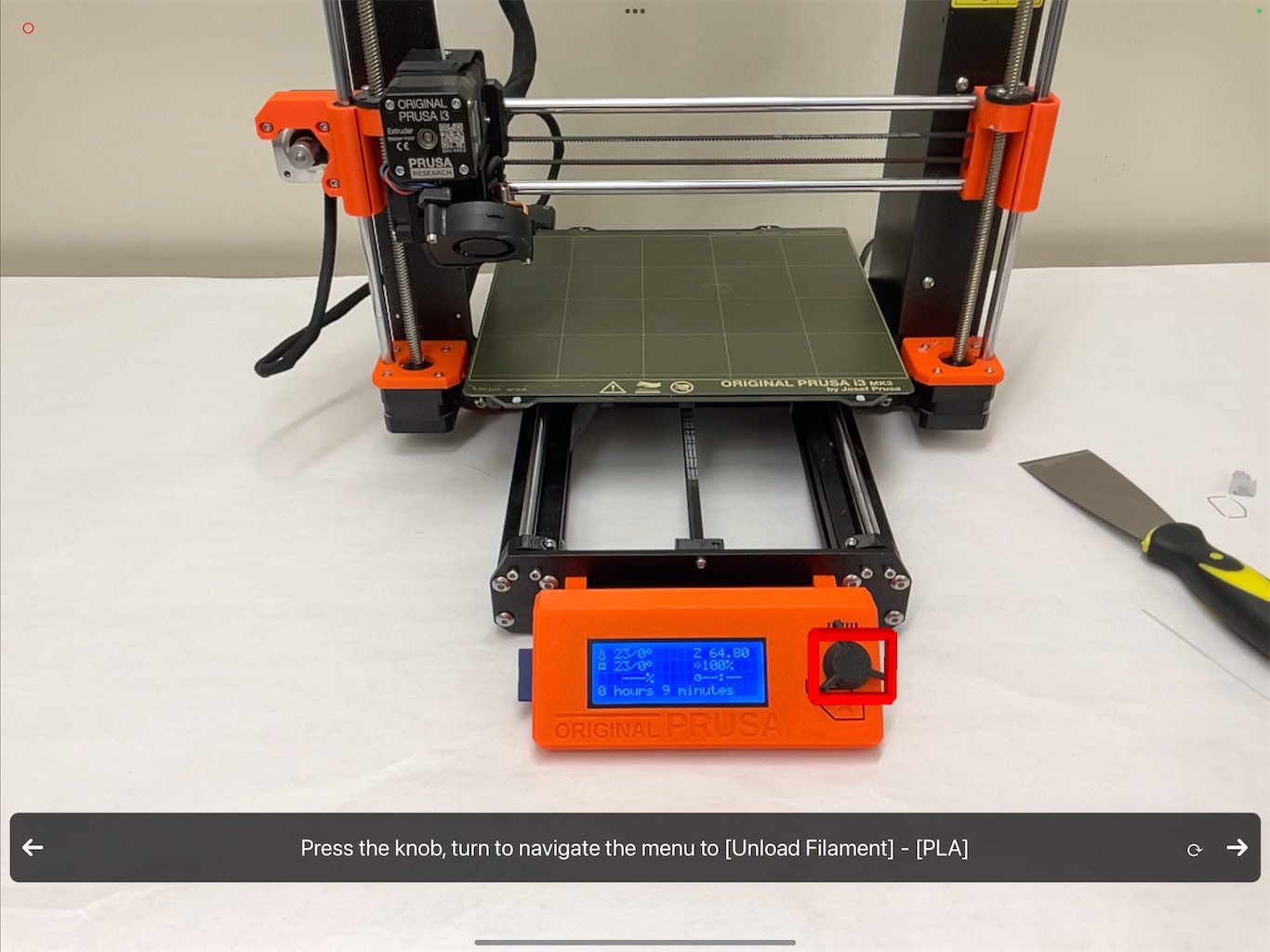}
\includegraphics[width=0.24\linewidth]{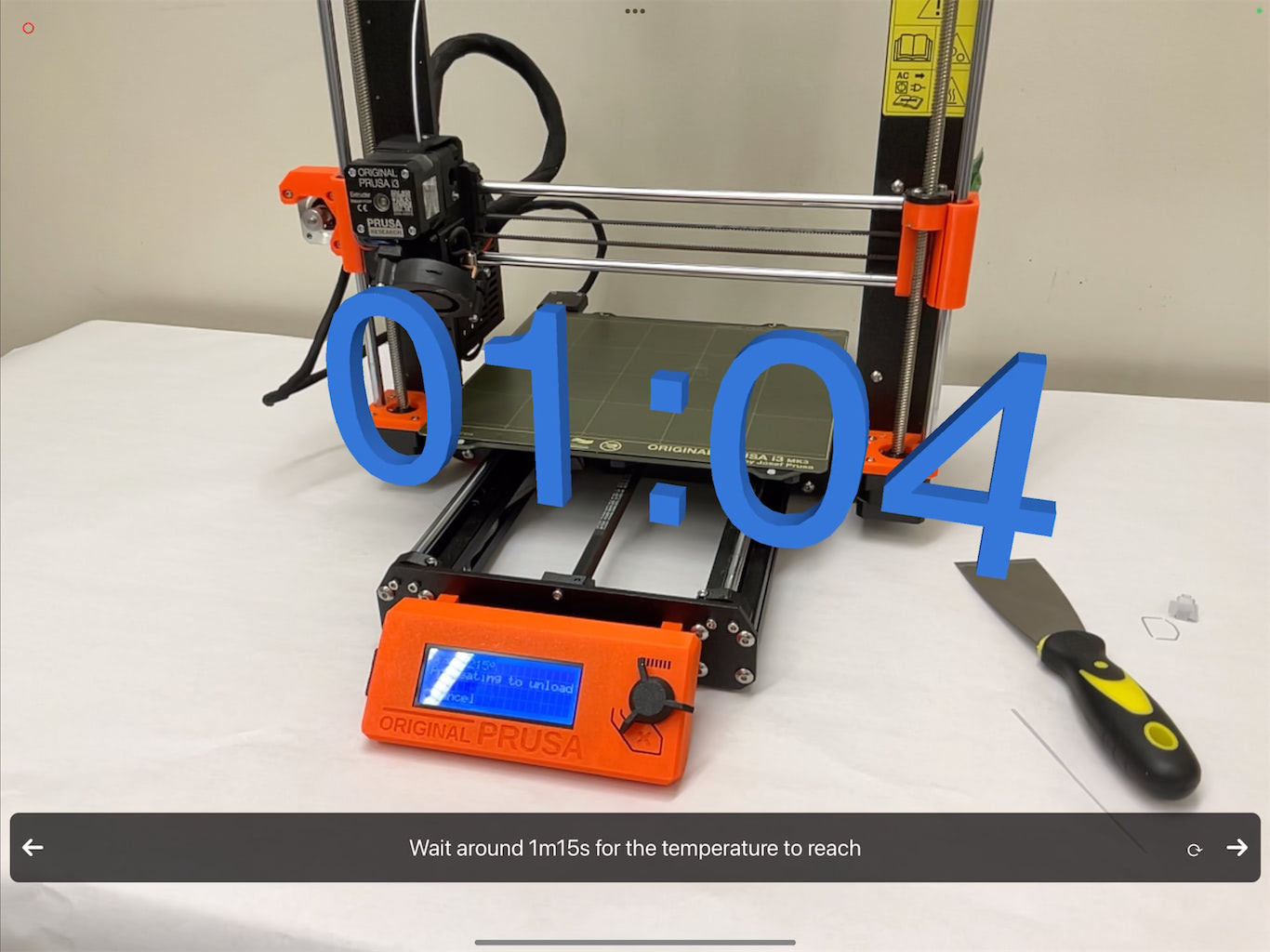}
\includegraphics[width=0.24\linewidth]{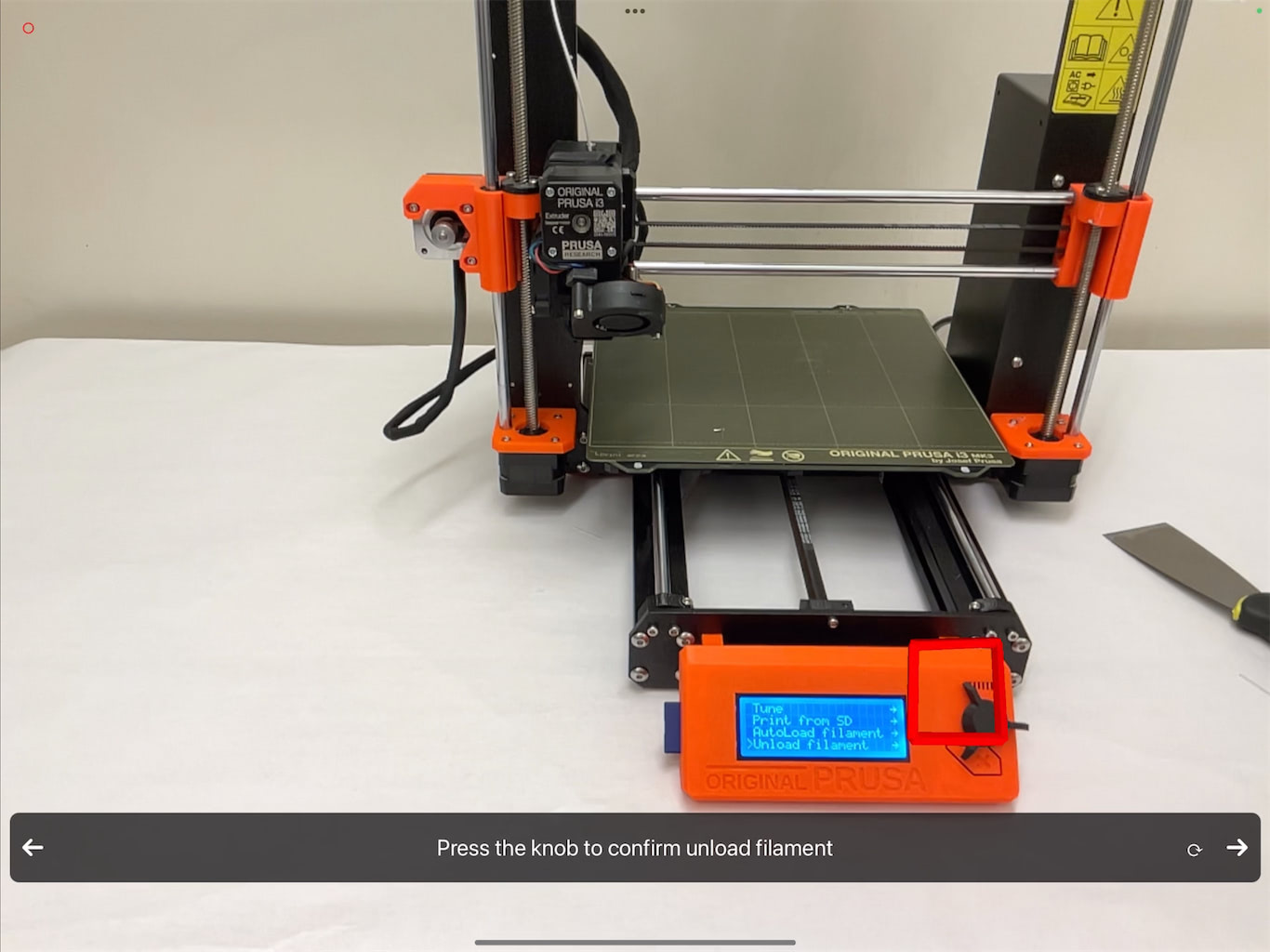}
\includegraphics[width=0.24\linewidth]{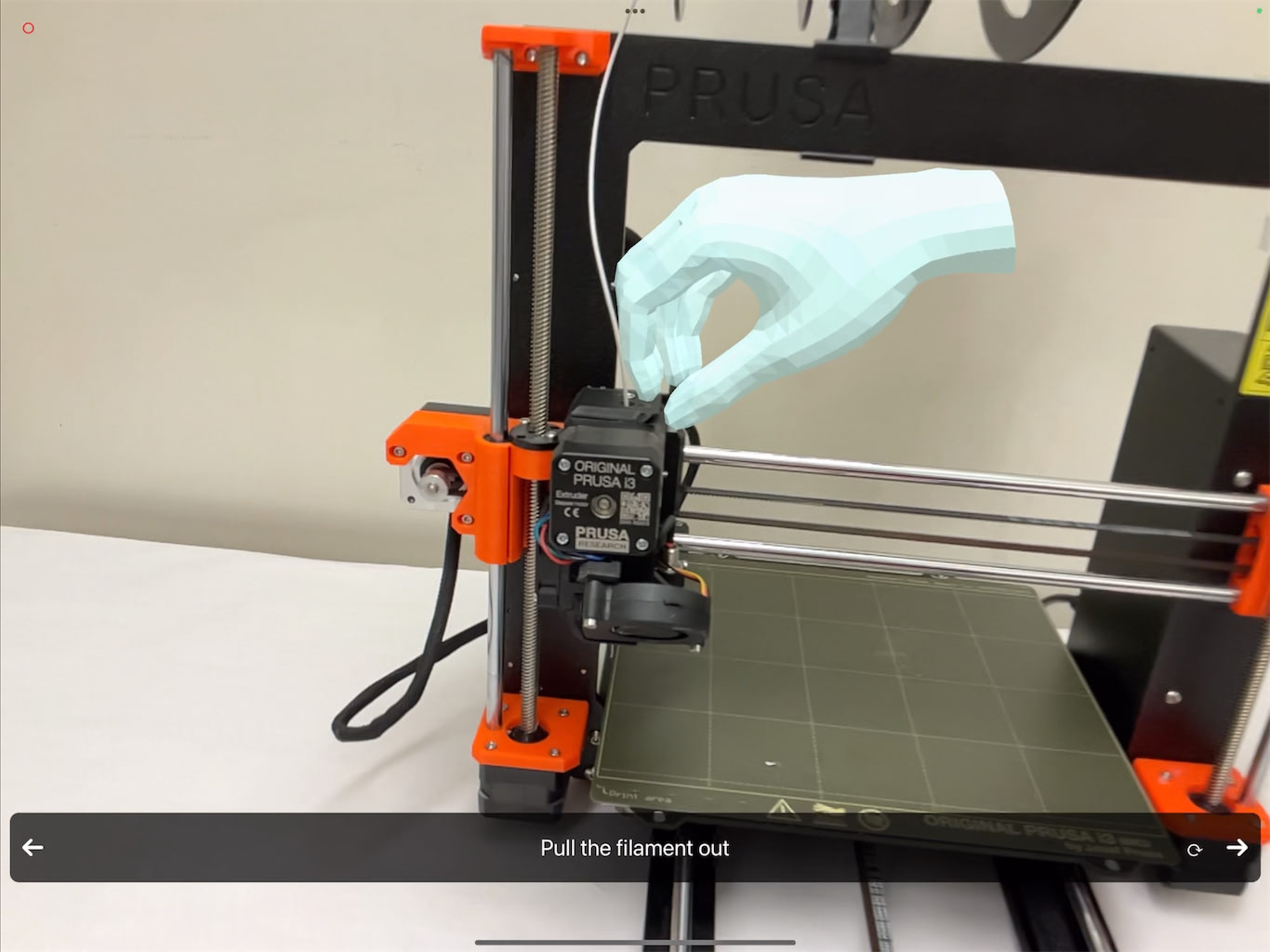}
\caption{The Guided Reality system provides visual step-by-step instructions to assist users in resetting a 3D printer. The image shows: (1) setup, (2) locating the 3D printer with a \textit{bouncing box highlight}, (3) scraping the printed object with an \textit{animated tool}, (4) moving the print bed back to the center with \textit{movement of a segmented object}, (5) selecting "unload filament" using the knob with a \textit{bounding box highlight}, (6) waiting 1 minute and 30 seconds for the nozzle to heat with a \textit{contextual timer widget}, (7) confirming unload using the knob with a \textit{bounding box highlight}, and (8) pulling the filament out with a \textit{hand gesture}.}
\label{fig:3d-printer}
\end{figure*}

\section{Guided Reality System Design}
We present Guided Reality, an AR instructional system that generates and embeds visually rich, spatially grounded task guidance. Given a user’s natural language question about a task, the system generates a step-by-step structured task plan, where each step includes both text and visual guidance embedded in the physical environment. We built our prototype on an iPad Pro to leverage camera access and real-time LiDAR depth. While not ideal for hands-free AR, it was the most practical option. Headsets like Meta Quest lacked camera access until March 2025, and HoloLens’s narrow field of view limited interaction. Though we plan to support head-mounted displays in the future, mobile devices offer greater mobility, accessibility, and scalability today, which enables us to test the system across diverse scenarios. The following section provides a detailed walkthrough of how the system responds to user input, processes the environment, and delivers interactive guidance throughout a multi-step task.

\subsection{System Walkthrough}
At the start of the experience, the user enters a task-related question into the text field at the top of the interface. For example, the user might type \textit{``how to clean the 3D printer from this stage''} (Figure~\ref{fig:3d-printer}). The system captures the camera view and takes the user’s question to query ChatGPT. We prompted the model to return a JSON data about a structured task plan where each step includes the textual instruction, the key component that the user needs to interact with, the appropriate visual guidance type, and any supporting metadata needed to generate that visual aid.

Once the structured task plan is retrieved, the system begins generating the appropriate visual guidance for the first step based on identified visual type. The user navigates through the steps using left and right arrow buttons on the side. Each time the user advances to a new step, the system repeats the visual guidance generation pipeline using the updated camera view. This visual guidance generation pipeline consists of three sub-processes.

\subsubsection*{\textbf{Generates Visual Content}} First, the system captures a new image of the scene and identifies the key component referred to in the instruction. Using a visual understanding model such as Gemini, the system locates the component’s bounding box, and if necessary, segments it using tools like Segment Anything~\cite{ravi2024sam} to isolate the region.

\subsubsection*{\textbf{Localizes Key Component in 3D Space}} Second, based on the 2D coordinates from the camera image and the device’s spatial sensing capabilities, including depth map, intrinsic parameters, and camera pose, the system computes the corresponding 3D position using raycasting. This projection ensures that the visual guidance can remain anchored in place even as the user moves the device.

\subsubsection*{\textbf{Embeds Visual Guidance in Space}} Finally, the system places and orients the visual guidance content within the physical environment. Depending on the visual type, different spatial information is used. For example, if the instruction involves overlaying a 3D model representing the tool on the surface of the object, the system estimates the surface normal from three localized points to determine the proper orientation. In another case, such as guiding a rotation movement, the system places an arrow in 3D space to indicate the direction of action by identifying the relevant axis of rotation and determining the appropriate direction along that axis. This real-time generation of spatial guidance at each step ensures that users receive contextual and accurate support during the task.

%% file: 5-implementation.tex
\begin{figure*}
\centering
\includegraphics[width=\linewidth]{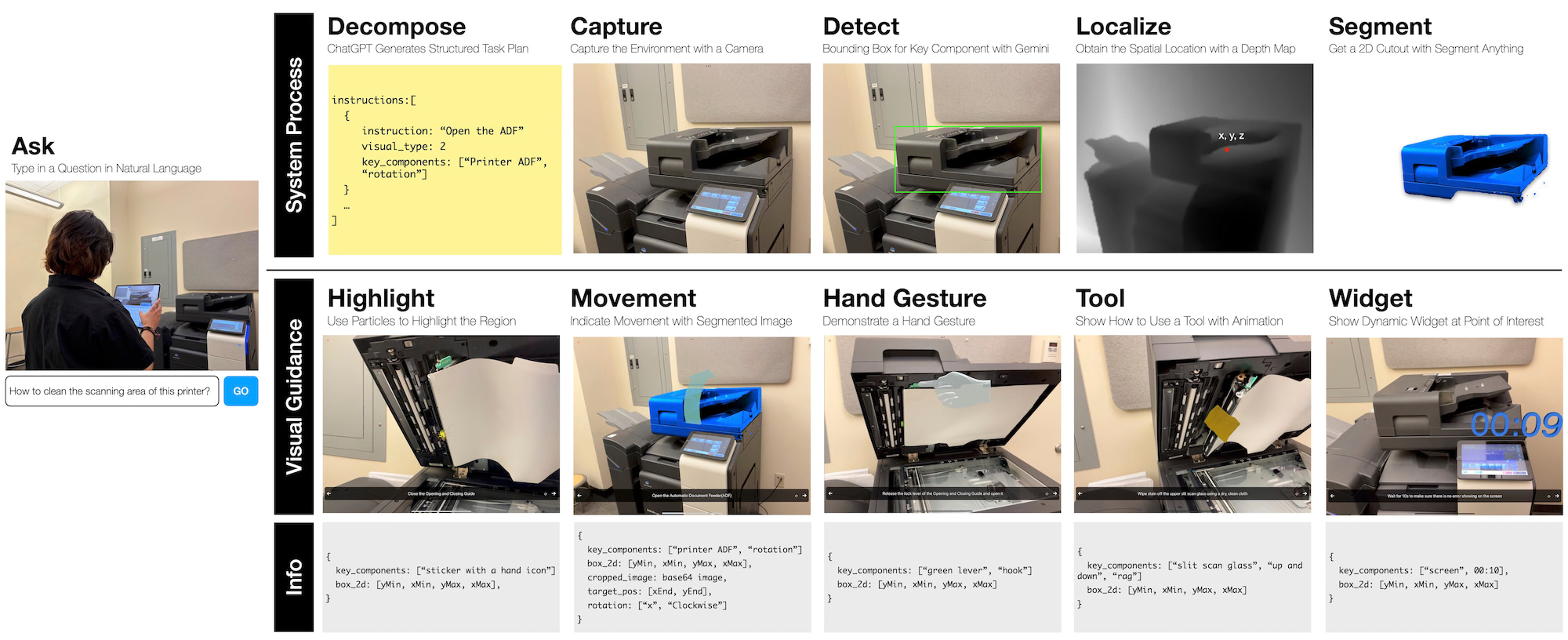}
\caption{System Implementation}
\label{fig:system}
\end{figure*}

\section{Implementation}

\subsection{Overall System Architecture}
The architecture of the Guided Reality system consists of two main components: a front-end iOS application developed in Swift and a back-end Python server running on a local PC. The front-end is responsible for user interaction, structured task plan generation, and AR content placement, while the back-end handles visual guidance generation. The iOS application runs on an iPad Pro (12.9-inch 6th Generation, Apple M2 Chip), utilizing UIKit for the user interface, ARKit for spatial tracking, and SceneKit for rendering 3D visual content. The back-end, running on a PC equipped with an NVIDIA GeForce RTX 2060 GPU, is dedicated to AI-based visual guidance generation for each instruction step.

When the user submits a natural language question along with a captured camera image, the front-end makes a call to ChatGPT API (gpt-4o model) to get a structured task plan. Each time the user advances to a new step, the iPad sends the current instruction step and an updated camera view to the back-end to request corresponding visual guidance content and placement data. Because visual guidance generation is computationally intensive, the iPad simultaneously stores the depth map and camera pose at the time of the request, enabling accurate spatial placement of the returned content within the AR scene.

For real-time communication, the iPad connects to the back-end using a WebSocket protocol to ensure high-resolution and low-latency performance. In scenarios where the iPad is not able to connect to the same local network—such as fixing car outside—a fallback mechanism using Firebase Realtime Database is employed to relay messages between the front-end and back-end.

The back-end is responsible for all AI-driven processing. Although large language model inference for instruction generation can be executed without a GPU, the back-end also runs vision models such as Segment Anything, which require GPU acceleration. To streamline the pipeline, all AI processing is centralized on the PC. Upon receiving a request, the back-end computes outputs including 2D bounding boxes of key components, and specific information according to the visual guidance type. These results are then transmitted back to the iPad. 

The front-end calculates final placement of the visual guidance using the previously saved depth map and camera pose. ARKit is used to anchor the guidance in 3D space, and SceneKit is used to render visual content within the augmented view. Figure~\ref{fig:system} illustrates the overall system implementation.

\subsection{Structured Task Plan Generation}
When the user initiates a question, we ask them to point the iPad straight to the device. The system relies on this initial image to establish spatial coordinates for conveying spatial information for visual guidance placement. The captured image is also used to identify the device brand to enable more specific instructions for visual guidance accurately. Using the user’s natural language question, the captured camera view, and a predefined prompt, the system queries ChatGPT-4o to decompose the request into a structured task plan in JSON format.

The predefined prompt plays a crucial role in ensuring consistent and interpretable outputs. It includes the objective of the system, an explanation of the visual guidance strategy, a detailed description of the expected JSON structure with example JSON snippets for each visual guidance type. The prompt is included in Appendix \ref{appendix:taskPlanPromp}.

The structured task plan contains an array, where each object specifies the instruction for a single step. Each step includes three fields: \texttt{instruction}, which contains the textual instruction; \texttt{visual\_type}, which specifies the appropriate visual guidance type based on our proposed strategy; and \texttt{key\_components}, a list that includes the description of the key component along with supplementary information that differs for each visual guidance type.

Because the description of the key component is used to interface with other AI models for generating visual guidance later on, we carefully evaluated the phrasing to optimize downstream performance. Preliminary testing showed that including additional contextual details, such as relative positions or visual attributes (e.g., colors, icons, shapes), improved object detection accuracy. Based on these findings, we explicitly included this in the predefined prompt so that the system incorporates such information in the \texttt{key\_component} field.

\subsection{Visual Guidance Generation}

When users progress to a new step, the system uses the updated camera view, and the current step information from the structured task plan to generate visual guidance. The visual type is determined by the \texttt{visual\_type} field.
In the following subsections, we describe each type of augmented visual guidance in detail.

\subsubsection*{\textbf{D1. Highlighting a Component}}
For region-of-interest visual guidance, the system captures an image from the updated camera view and stores the current camera pose, the LiDAR depth map. The back-end then uses the captured image and the first object in the \texttt{key\_component} to query Google Gemini 2.5 with the prompt in Appendix \ref{appendix:boxPromp} to retrieve a 2D bounding box of the key component. Compared to traditional object detection models like YOLO\cite{redmon2016you}, Gemini demonstrates better performance in identifying detailed components within complex devices.

Once the 2D bounding box is returned, the front-end retrieves the depth value at each corner of the box from the previously saved depth map. Using the camera’s intrinsic parameters, it projects the 2D corner coordinates back into 3D space relative to the saved camera pose. The resulting 3D corners are then connected to form a bounding box in space, visually highlighting the component the user needs to interact with. To improve visibility and reduce visual clutter, the system adapts the rendering strategy based on the physical size of the bounding box. When any edge of the box is shorter than 5 cm in 3D space, the bounding box is replaced with a particle effect located at its center.

\begin{figure}[h]
\centering
\includegraphics[width=0.49\linewidth]{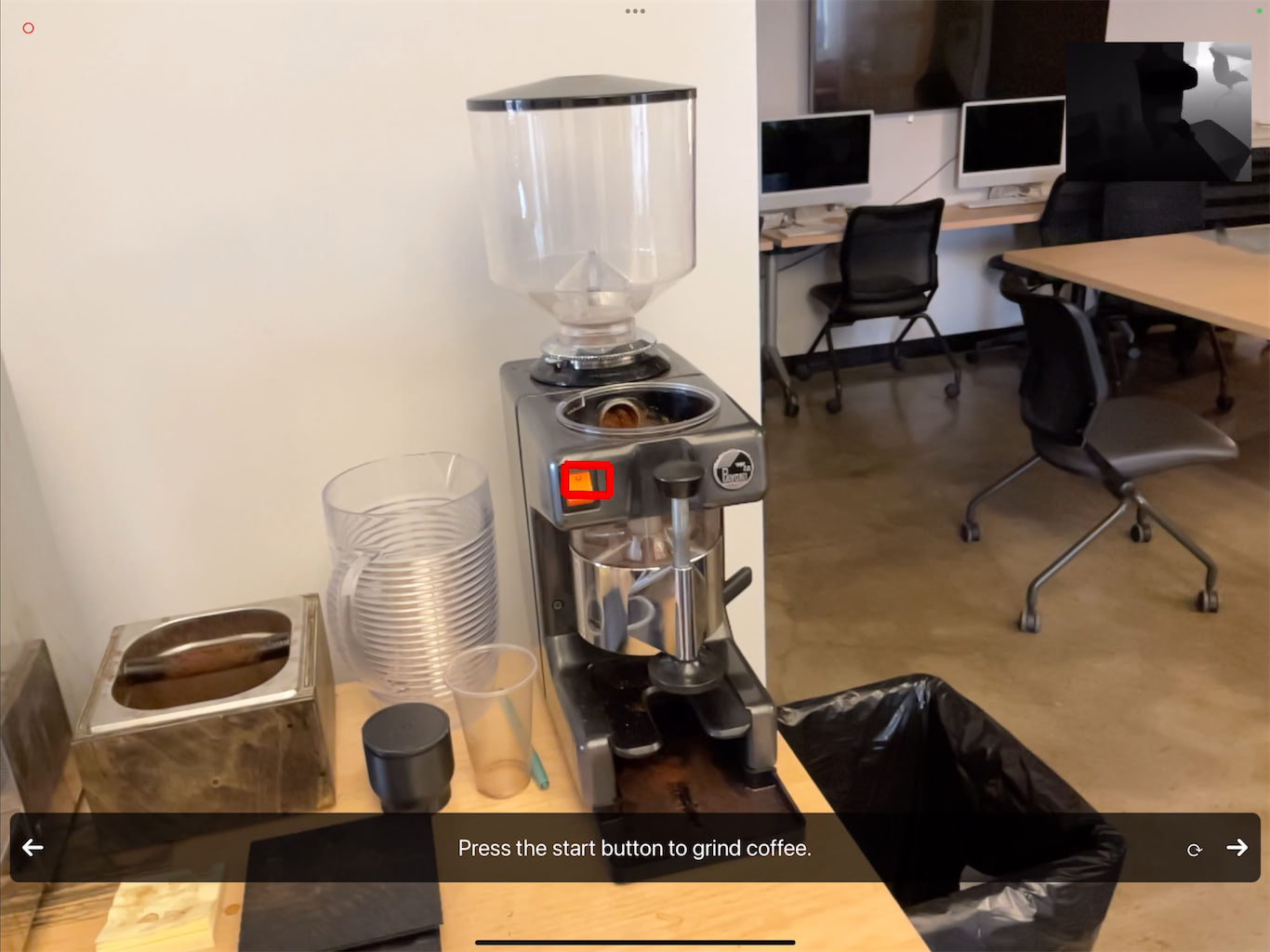}
\includegraphics[width=0.49\linewidth]{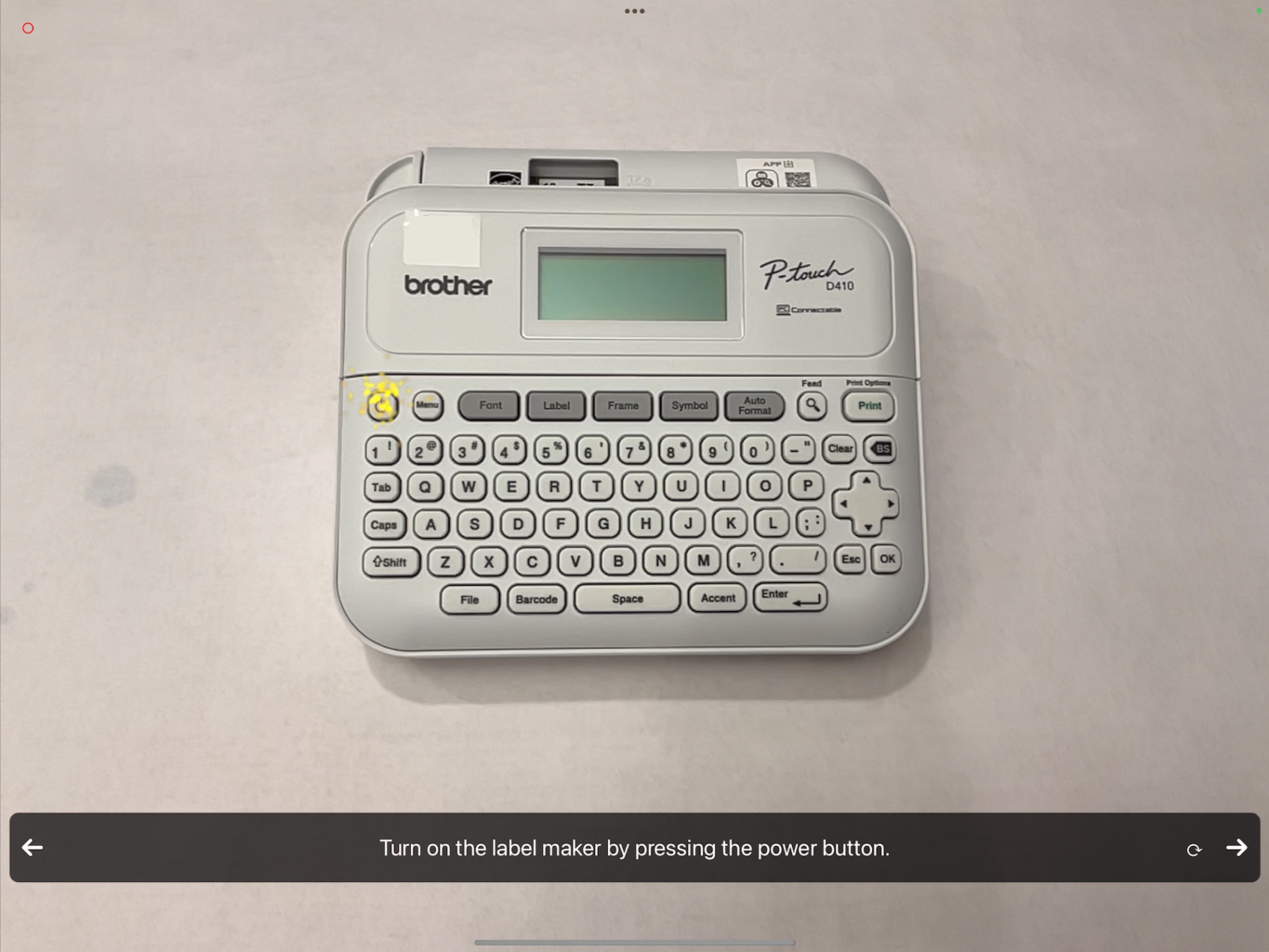}
\caption{Highlighting a Component}
\Description{A diagram showing the placement of arrows for rotational movement guidance.}
\label{fig:region}
\end{figure}

\subsubsection*{\textbf{D2. Indicating Movement}}
When the action involves movement, the system highlights the component to be moved in blue. For translational movements, the system animates the transition of the component from its original location to its target location. In the case of rotational movement, an arrow is rendered to indicate the movement. The type of movement is included in the original structured task plan.

\vspace{0.2cm} \noindent
\textbf{Translational Movement:} For movements such as “Move the printer bed back to the center,” we use a dedicated prompt, as shown in Appendix~\ref{appendix:translationalPromp} to query Gemini, requesting both the 2D bounding box of the key component and its estimated target position given the instruction. Using the identified bounding box, we apply Segment Anything 2.1~\cite{ravi2024sam} to segment the component in 2D. The segmented region is then cropped, and its blue channel is enhanced threefold using OpenCV and NumPy. The bounding box information, target position and cropped image are then returned to the front-end for further processing.

To display the moving component in 3D, we place the cropped image on an image plane. The scale of this image plane is calculated using the camera’s focal length and the depth value at the center of the bounding box. The image plane is then rotated to face the previously saved camera pose so that the highlighted component appears correctly situated in the physical environment, even though it is rendered as a 2D image overlay.
% \[
% \text{scaleX} = \left(\frac{\text{cropped\_image\_width}}{\text{camera\_focal\_length\_x}}\right) \times \text{depth\_at\_center}
% \]
We localize both the original and target positions of the component in 3D with the depth map. An animation is created by interpolating between these two 3D points, guiding the user’s movement of the component.
\begin{figure}[h]
\centering
\includegraphics[width=0.49\linewidth]{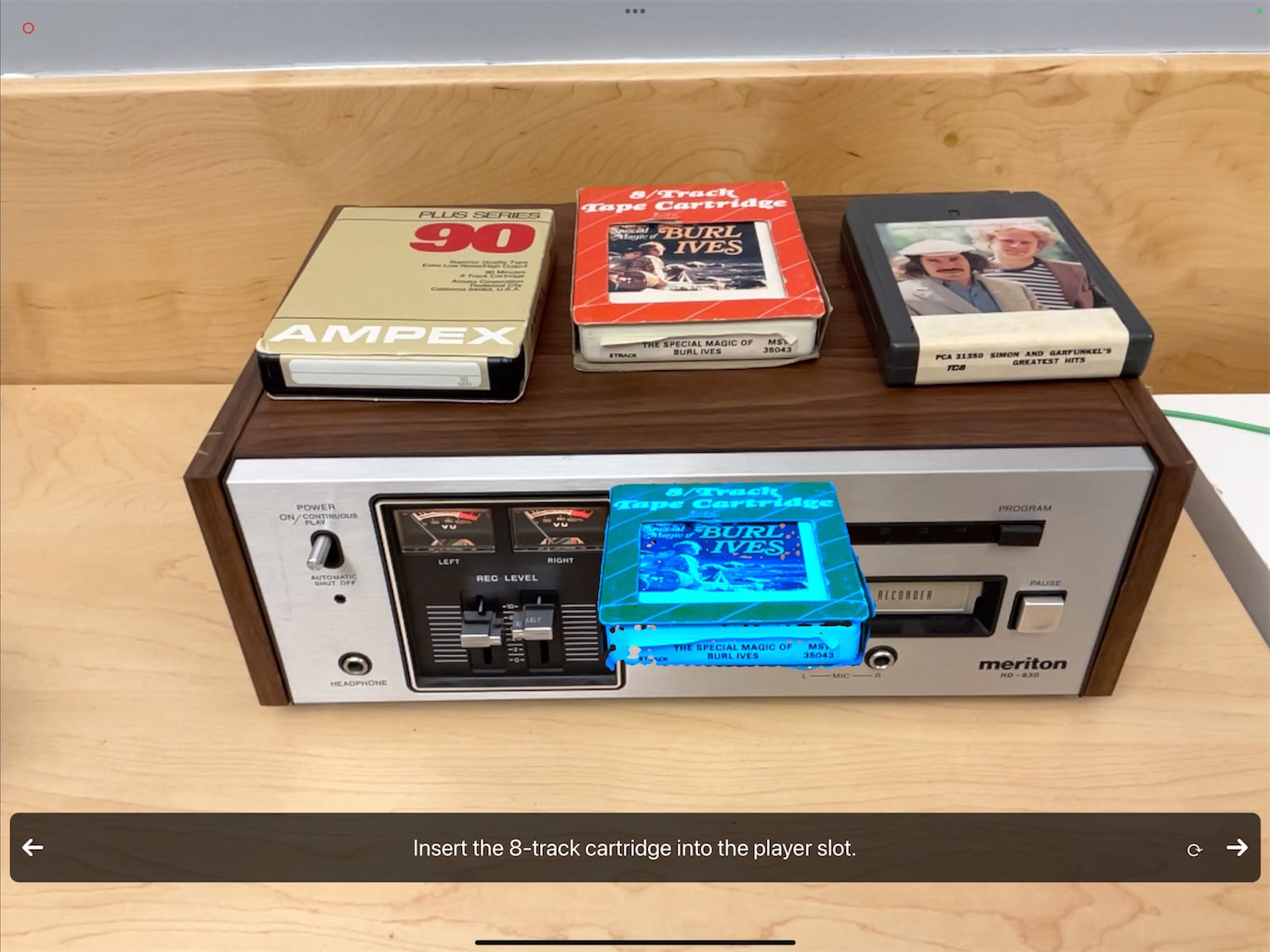}
\includegraphics[width=0.49\linewidth]{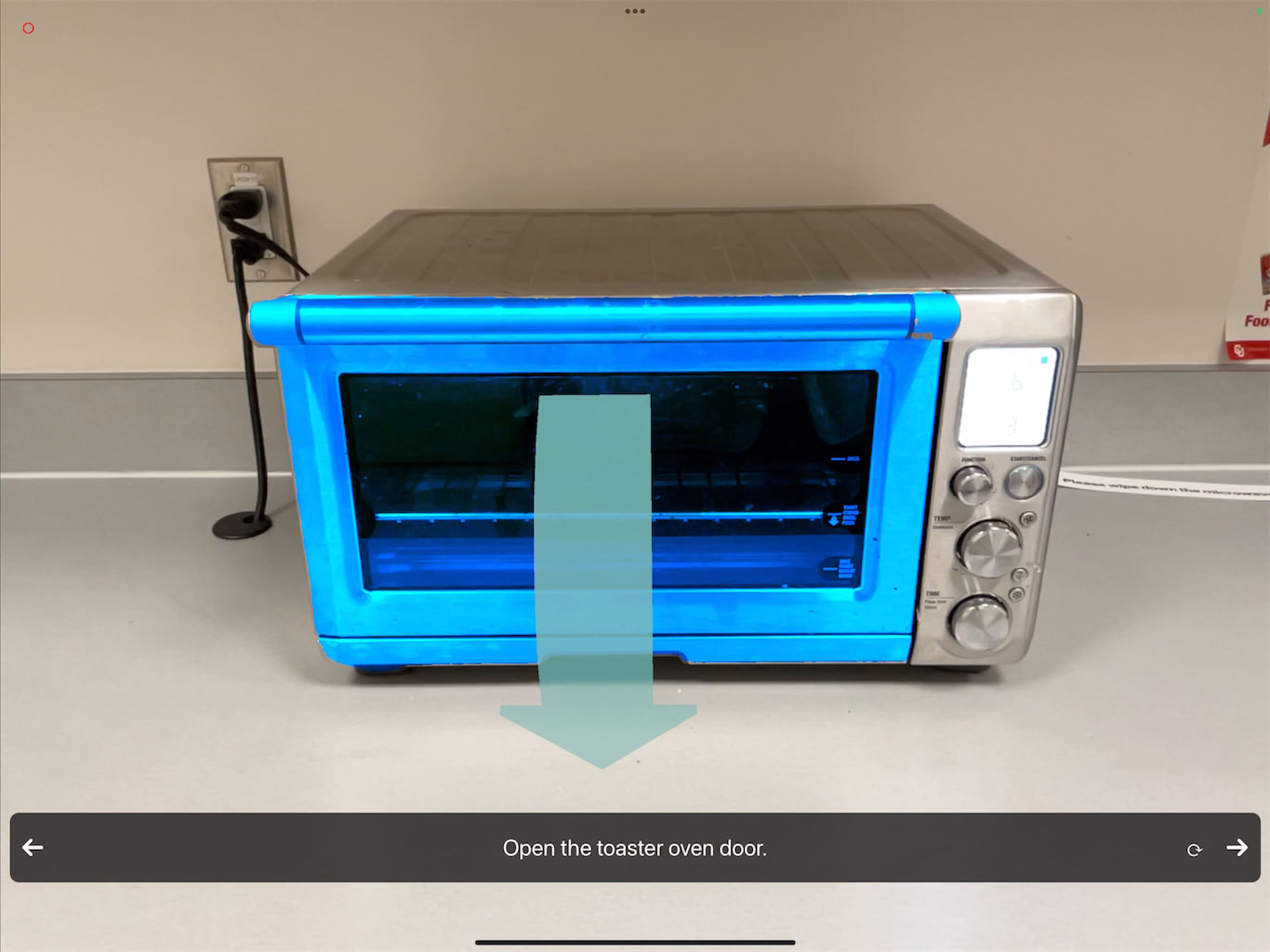}
\caption{Indicating Movement}
\Description{A diagram showing the placement of arrows for rotational movement guidance.}
\label{fig:movement}
\end{figure}

\vspace{0.2cm} \noindent
\textbf{Rotational Movement:} For actions involving rotation, the system highlights the relevant component and overlays a curved arrow to show the direction. The system uses the initial image submitted with the user’s question, along with the current camera view, to query ChatGPT 4o for rotation details using the prompt provided in Appendix~\ref{appendix:rotationalPrompt}. The initial image set the spatial axis with x pointing right, while the current view shows the current status. To standardize how rotations are described, we define: (1) the axis of rotation and (2) the direction when viewed from the positive side of that axis. For example, to open a toaster oven (Figure~\ref{fig:movement}.2), the system identifies the \textit{x}-axis and a counterclockwise rotation. The curved arrow is centered on the key component, scaled to match its size, and aligned with the identified axis and direction.

\subsubsection*{\textbf{D3. Demonstrating a Hand Gesture}}
The current system uses six static hand poses inspired by Cutkosky’s grasp taxonomy~\cite{cutkosky1989grasp}, representing common one-handed interactions in machine operation tasks. These include: poke, hook, palm press, grip, cylindrical grasp, and pinch. Each gesture model is pre-aligned with a common orientation, where the typical point of contact is positioned at the origin. When the structured task plan identifies hand gesture as the visual type, it also selects the most appropriate pose from this predefined set. To place the hand in the AR scene, the system projects the center of the 2D bounding box of the key component into 3D space and positions the gesture model accordingly, oriented to face the saved camera pose.

\begin{figure}[h]
\centering
\includegraphics[width=0.49\linewidth]{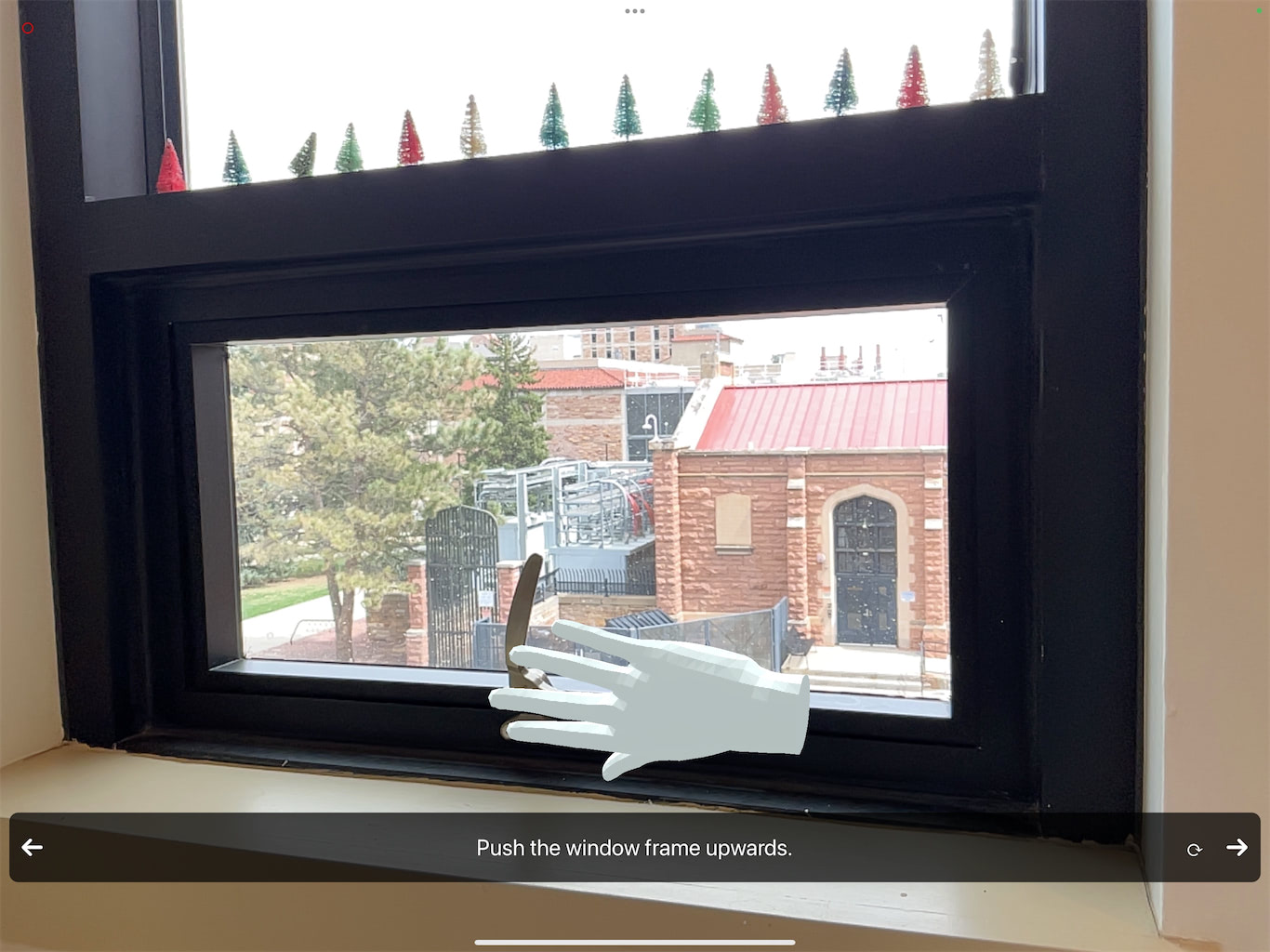}
\includegraphics[width=0.49\linewidth]{figures/printer-11.jpg}
\caption{Demonstrating a Hand Gesture}
\Description{A diagram showing the placement of arrows for rotational movement guidance.}
\label{fig:region}
\end{figure}

\subsubsection*{\textbf{D4. Showing How to Use a Tool}}
When an instruction involves using a tool, the system displays a 3D model positioned and oriented in space to convey what the tool is, where to use it, and how. This guidance depends on both the model’s geometric accuracy and correct spatial placement. We experimented with generation tools such as Meshy~\cite{MeshyAI_Text_to_3D} and Shap-E~\cite{jun2023shap}, but found that the outputs often lacked detail and were frequently misoriented, even when asked to align the functional end with the XZ surface. Thus, we curated a library of commonly used tools with the correct scale and orientation, while using Meshy~\cite{MeshyAI_Text_to_3D} as a fallback for tools not in the collection.

To determine spatial placement, the system first identifies the 2D bounding box of the key component. It then projects three points, bottom-left, bottom-right, and center, into 3D space to compute the surface normal, which represents the orientation of the contact surface. The tool is positioned at the 3D point projected from the center of the 2D bounding box, with its local \textit{y}-axis (up direction) aligned to the surface normal. To face the interaction direction, the tool is further rotated to point toward the projected bottom-center point. This approach ensures that the tool appears naturally aligned with the target surface in the AR scene.

We support two types of movement indications for tool-related actions: translational animation and self-rotation using a curved arrow. The structured task plan specifies the movement type—either “up and down,” “left and right,”,  “rotation”, "clockwise" or "counter clockwise". For translational actions, the tool is animated between points projected from the bounding box such as top-center to bottom-center for vertical motion. For self-rotational actions, like tightening or loosening a screw, The system then renders a curved arrow by applying the tool’s transformation matrix and rotating it ±90° along its local \textit{x}-axis, visualizing the intended motion in context.

\begin{figure}[h]
\centering
\includegraphics[width=0.49\linewidth]{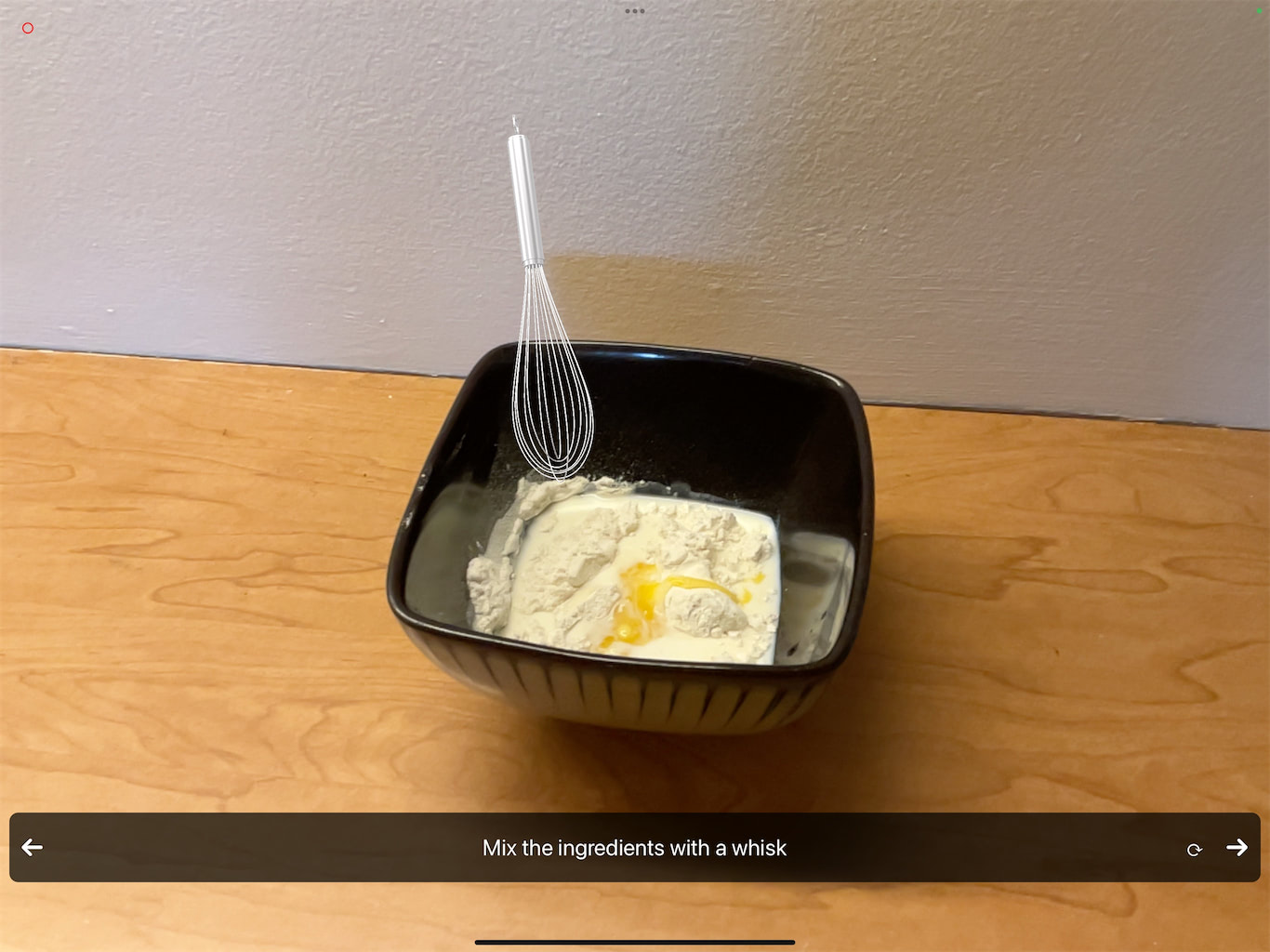}
\includegraphics[width=0.49\linewidth]{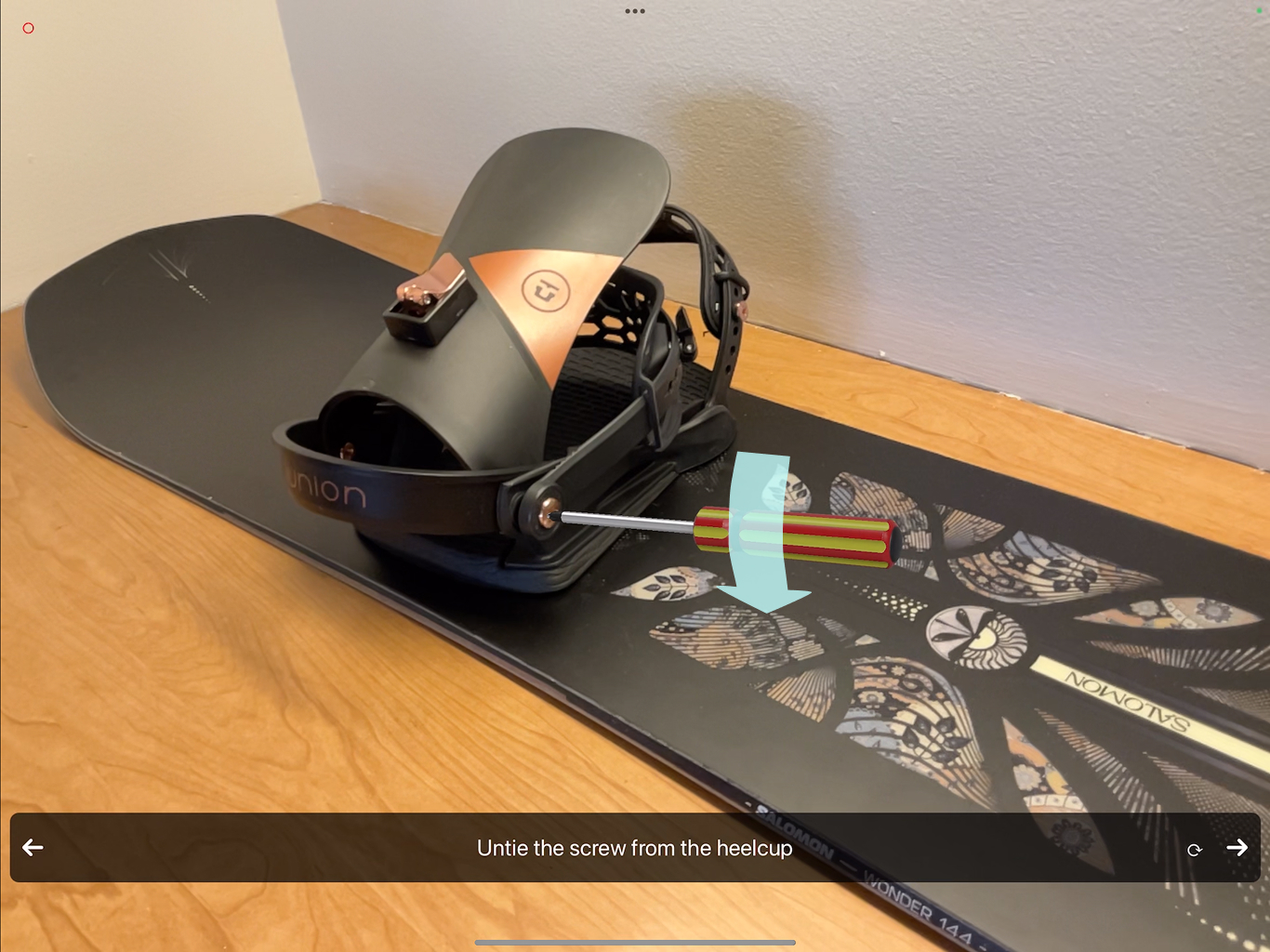}
\caption{Showing How to Use a Tool}
\Description{Showing How to Use a Tool}
\label{fig:region}
\end{figure}

\subsubsection*{\textbf{D5. Displaying a Contextual Widget and Information}}
When users are waiting for a period during the task execution, we provide a contextual widget, a timer to help users keep track of the time. If the initial structured task plan has a waiting step, it also outputs the estimated waiting time. By localizing points in 3D with 2D bounding box and depth map, we display a timer at the top-center of the identified key component, facing the saved camera pose.

% \begin{figure}[h]
% \centering
% \includegraphics[width=0.49\linewidth]{figures/laminator-1.jpg}
% \includegraphics[width=0.49\linewidth]{figures/oven-2.jpg}
% \caption{Displaying a Contextual Widget and Information}
% \Description{A diagram showing the placement of arrows for rotational movement guidance.}
% \label{fig:region}
% \end{figure}

\subsection{Technical Evaluation}
\subsubsection{Methods}

\paragraph{User Query–Driven Evaluation.}
To evaluate the accuracy and versatility of our generative instruction pipeline in response to a user query, we conducted a technical evaluation with three tasks from each domain listed in Section~\ref{sec:task_domains}. For each task, we applied our system using a standardized query in the form \textit{“How to [do something]”}, such as \textit{“How to make a cup of coffee?”} and \textit{“How to change the air filter?”}, and then pointed the system to the device. In total, we gathered 15 queries with 100 steps as our evaluation dataset. For each step, we evaluated whether the structured task plan information was correct, including the text instruction, the identified visual type and key component. If the structured task plan was accurate, we assessed the correctness of the final visual guidance and computed accuracy metrics for each visual type.

% Since bounding box detection is a shared process across all visual types, we assessed its accuracy across all steps. We then evaluated each type of visual guidance individually, based on its specific breakdown components.

\paragraph{Visual Type Evaluation.}
To mitigate the uneven distribution of visual types in user-driven queries, we conducted a second evaluation using a balanced dataset with 20 steps per visual guidance type. This evaluation measures how accurately each type can be embedded in space with a correct structured task plan. In addition, we assess component-wise accuracy and latency for each type.

\subsubsection{Results}

\paragraph{User Query–Driven Result.}
Table~\ref{tab:tech-evaluation} presents a summary of our user query-driven evaluation results. In general, the system produced the correct results for 65 out of 100 steps of the 15 tasks. Among the errors, 17 steps failed due to incorrect structured task plan generation, while 18 steps failed during the visual guidance generation and placement process. A key issue lies in visual type identification, which achieved 90\% accuracy under a strict classification scheme. We deliberately applied conservative labels during evaluation, marking a visual type as incorrect even if the system-generated type could reasonably support user understanding. For example, the instruction “hold the paper in place using the plastic handle on top” was ideally labeled as \textit{Movement}, but the system identified it as \textit{Highlight}. While this was marked as a misclassification, the resulting guidance still effectively conveyed the intended action to the user. We will report the accuracy for each visual type together with the second evaluation in the next paragraph.

\paragraph{Visual Type Evaluation.}

Table~\ref{tab:type-evaluation} provides an overview of the visual type evaluation, including both accuracy and latency metrics. \textit{Highlight} steps struggled with identifying small or visually similar components—for example, locating the middle C key on a piano.
\textit{Translational Movement} steps achieved an 80\% accuracy rate. While the Segment Anything model~\cite{ravi2024sam} performed well overall, errors typically arose from incorrect identification of the target position. \textit{Rotational Movement} steps had a 70\% accuracy rate. All errors were due to misidentifying the rotation direction, with the correct rotational axis detected.
\textit{Hand Gesture} steps reached 75\% accuracy. Failures were often caused by placement errors when gestures involved non-planar contact points, such as gripping a drill handle. Our raycasting pipeline occasionally overshoots the handle and intersects with the background, resulting in incorrect placement with an inaccurate surface normal.
\textit{Tool} steps encountered difficulties in selecting small or visually similar components, such as screws on a bed frame. It also suffered from misalignment when the 3D model generation model failed to orient the tool correctly.
All \textit{Widget} steps were generated correctly, as they typically rely on bounding box detection of relatively large objects like a laminator.

In terms of system latency, most steps completed within five seconds. The exception occurred when the required external tool was missing from the library, triggering the 3D model generation pipeline and increasing latency.

\renewcommand{\arraystretch}{1.0}
\begin{table}[h]
\centering
\begin{tabular}{@{}l c c c@{}}
\toprule
\textbf{} & \textbf{Total Steps} & \textbf{Correct Steps} & \textbf{Percentage} \\
\midrule
Text Instruction & 100 & 96 & 96.0\% \\
Visual Type      & 100 & 90 & 90.0\% \\
Key Component    & 100 & 97 & 97.0\% \\
% \midrule
% Detection        & 83  & 71 & 84.2\% \\
\midrule
Highlight        & 40  & 32 & 80.0\% \\
Movement         & 20  & 17 & 85.0\% \\
Hand Gesture     & 14  & 11 & 78.6\% \\
Tool             & 4   & 3  & 75.0\% \\
Widget           & 5   & 5  & 100.0\% \\
\midrule
\textbf{Total}   & 100 & 65 & 65.0\% \\
\bottomrule
\end{tabular}
\vspace{0.5em}
\caption{Technical evaluation results across 100 steps from 15 tasks with user queries.}
\label{tab:tech-evaluation}
\end{table}

\renewcommand{\arraystretch}{1.0}
\begin{table}[h]
\centering
\begin{tabular}{@{}l c c@{}}
\toprule
\textbf{Type / Component} & \textbf{Accuracy} & \textbf{Latency (s)} \\
\midrule
\textbf{Highlight}              & 90\%  & 3.29 \\
\quad 2D Box                    & 90\%  & 3.23 \\
\midrule
\textbf{Translational Movement} & 80\%  & 4.03 \\
\quad 2D Box                    & 100\% & 3.49 \\
\quad End Position              & 80\%  & - \\
\quad Segmentation              & 90\%  & 0.46 \\
\midrule
\textbf{Rotational Movement}    & 70\%  & 4.09 \\
\quad 2D Box                    & 100\% & 3.36 \\
\quad Rotation Info             & 70\%  & 2.41 \\
\quad Segmentation              & 100\% & 0.47 \\
\midrule
\textbf{Hand Gesture}           & 75\%  & 3.31 \\
\quad 2D Box                    & 100\% & 3.25 \\
\quad Type                      & 90\%  & – \\
\quad Placement                 & 85\%  & – \\
\midrule
\textbf{Tool}                   & 75\%  & 3.29 (29.23 w/ gen) \\
\quad 2D Box                    & 80\%  & 3.24 \\
\quad Tool Gen (N=3)            & 33\%  & 23.80 \\
\midrule
\textbf{Widget}                 & 100\% & 3.30 \\
\quad 2D Box                    & 100\% & 3.24 \\
\bottomrule
\end{tabular}
\vspace{0.5em}
\caption{Per-type accuracy and latency for visual guidance and component breakdown.}
\label{tab:type-evaluation}
\end{table}

%% file: 6-user-study.tex
\section{User Study}
\begin{figure*}
\centering

\label{fig:demo-3d-printer} % \label{fig:userstudy}
\end{figure*}
We evaluate our system with a user study and an instructor interview to gain feedback from two sides. 

\subsection{Methods}
The first study's goal is to evaluate the usability of our system and the effectiveness of visual guidance. We recruited 16 participants (8 female, 8 male, age 21-30) from our local community. This user study consists of the following two phases:

\subsubsection{Designated Tasks}

At the beginning of the session, participants were welcomed and given an overview of the user study. We walked them through the user interface and clarified that the system has limitation for hands-free operations so participants could hand the iPad to the researcher at any point.

The study employed a within-subjects design. Each participant completed two tasks: (1) cleaning the scanning area of a Konica Minolta bizhub C450i (Figure~\ref{fig:teaser}) and (2) resetting a Prusa I3 Mk3 3D printer (Figure~\ref{fig:3d-printer}). Both tasks had seven steps and included all five visual guidance types. To ensure consistency between participants, the same pre-generated structured task plan was used in each session for each task. Each participant completed one task using Guided Reality and the other using a baseline condition, where a static yellow arrow pointed to the key component. The order of task-condition pairing was counterbalanced. We collected participants’ task completion times and captured their operations through screen recordings during the user study. Participants completed a survey after each condition, evaluating their prior familiarity with the task and the system's usability using the System Usability Scale (SUS)~\cite{lewis2018system}. Additionally, they rated the effectiveness of each visual guidance type, which was collected only when using Guided Reality. Finally, they reviewed video recordings of their task performance and participated in a semi-structured interview to reflect on the intuitiveness, effectiveness, and engagement of the visual guidance compared to the baseline. Each session lasted approximately 30 minutes.
% \begin{figure*}
% \centering
% \includegraphics[width=0.24\linewidth]{figures/turntable-1.jpg}
% \includegraphics[width=0.24\linewidth]{figures/fire-extinguisher-1.jpg}
% \includegraphics[width=0.24\linewidth]{figures/cpr-1.jpg}
% \includegraphics[width=0.24\linewidth]{figures/cat-1.jpg}
% \\[0.05cm]
% \includegraphics[width=0.24\linewidth]{figures/lego-1.jpg}
% \includegraphics[width=0.24\linewidth]{figures/hammer-1.jpg}
% % \includegraphics[width=0.24\linewidth]{figures/desktop-printer-1.jpg}
% \includegraphics[width=0.24\linewidth]{figures/piano-1.jpg}
% \includegraphics[width=0.24\linewidth]{figures/slicer-1.jpg}
% \\[0.05cm]
% \includegraphics[width=0.24\linewidth]{figures/coffee-2.jpg}
% \includegraphics[width=0.24\linewidth]{figures/fire-extinguisher-2.jpg}
% \includegraphics[width=0.24\linewidth]{figures/software-1.jpg}
% \includegraphics[width=0.24\linewidth]{figures/tripod-1.jpg}
% \caption{\todo{in-the-wild user study}}
% \label{fig:in-the-wild}
% \end{figure*}

\begin{figure*}
\centering
\includegraphics[width=0.19\linewidth]{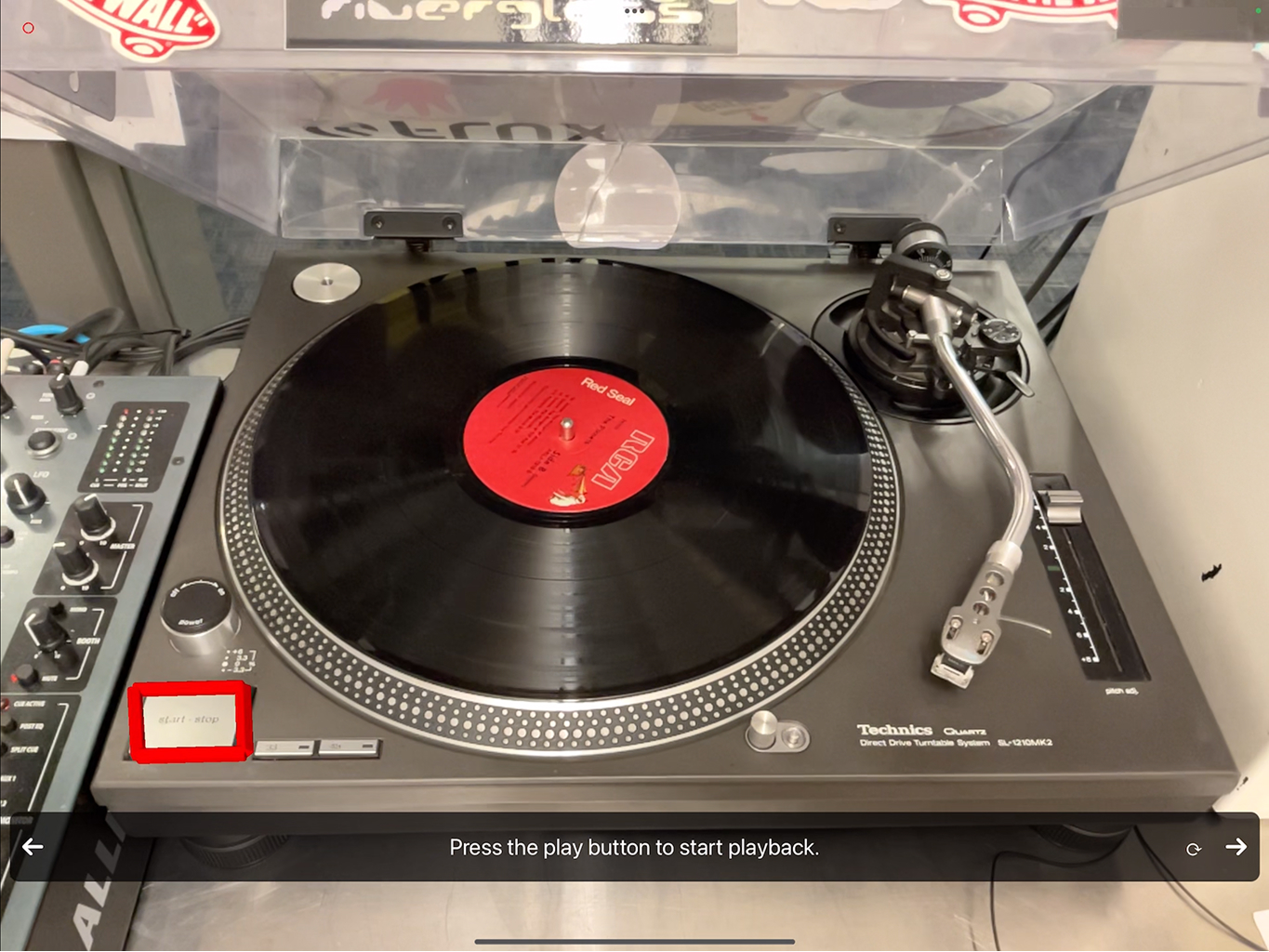}
\includegraphics[width=0.19\linewidth]{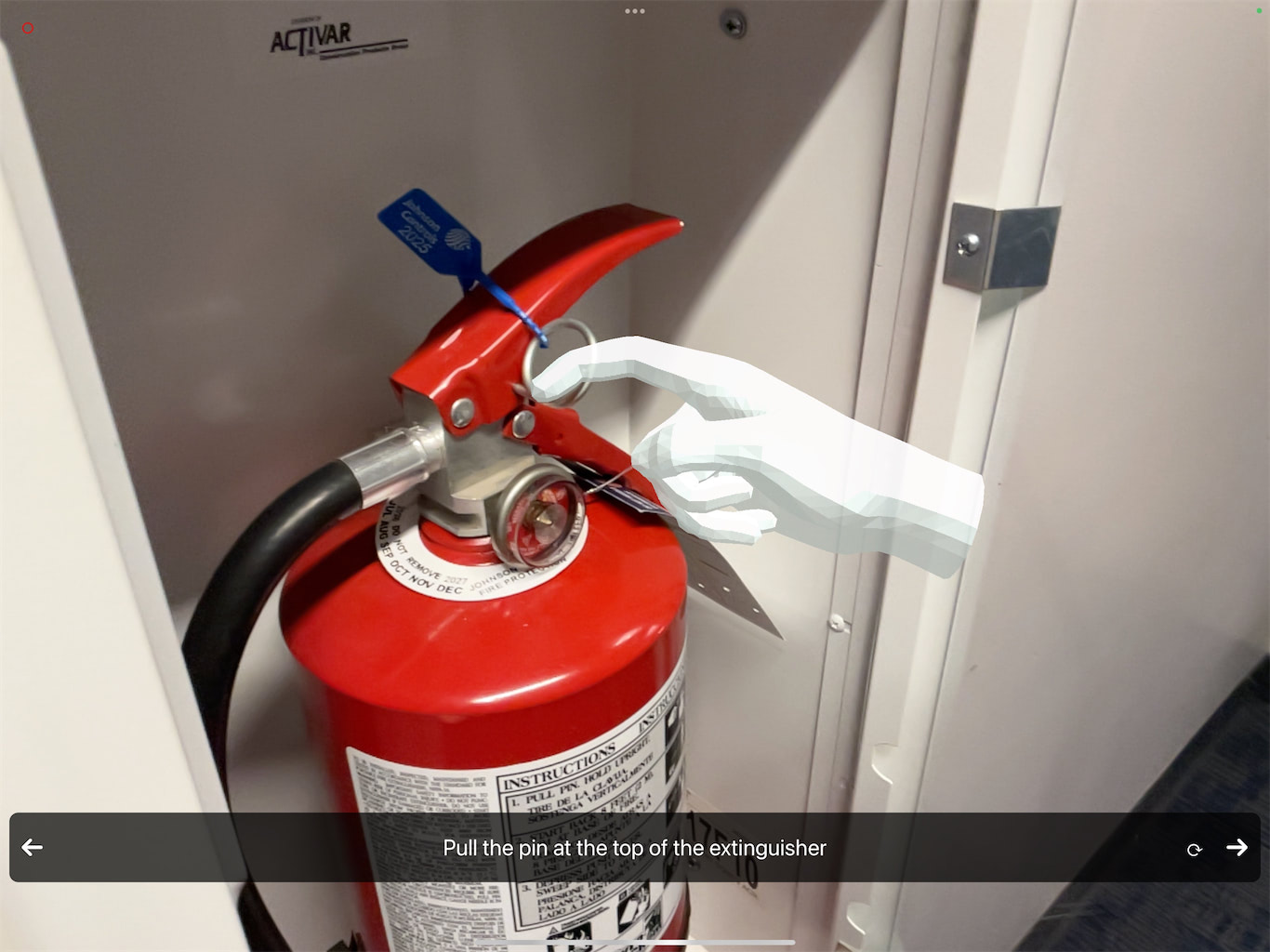}
\includegraphics[width=0.19\linewidth]{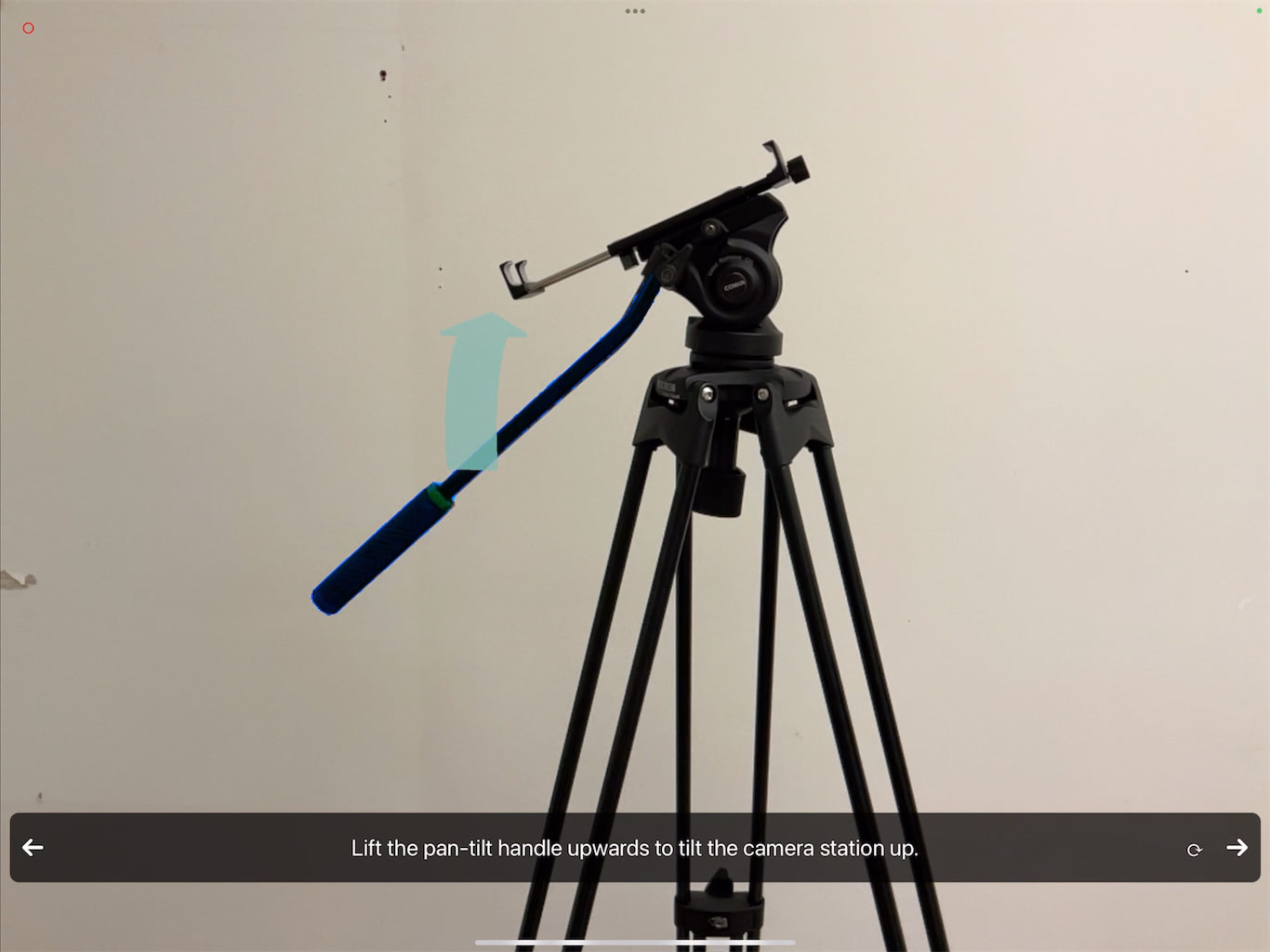}
\includegraphics[width=0.19\linewidth]{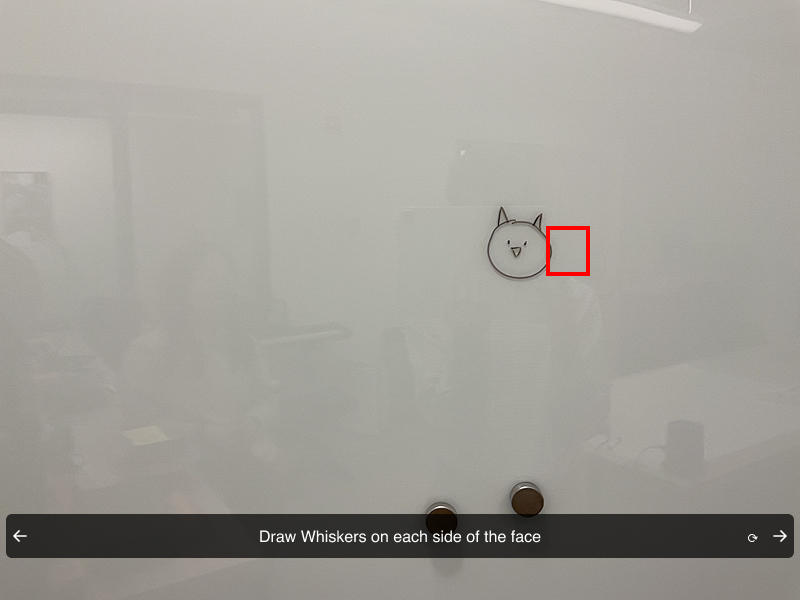}
\includegraphics[width=0.19\linewidth]{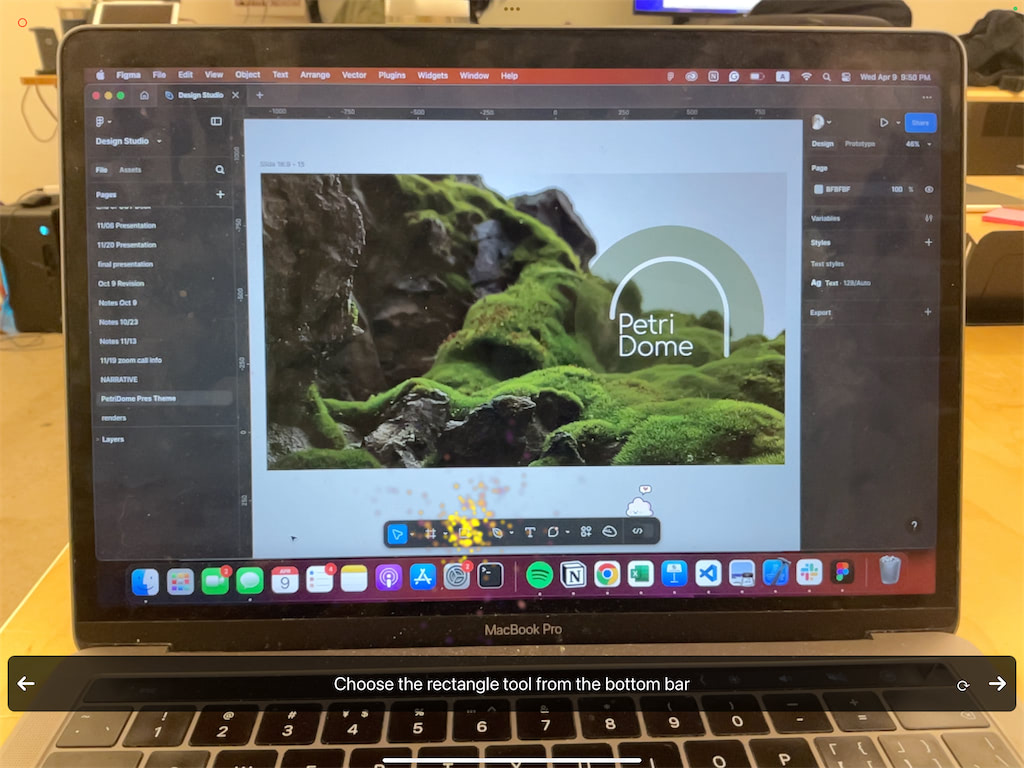}
\\[0.05cm]
\includegraphics[width=0.19\linewidth]{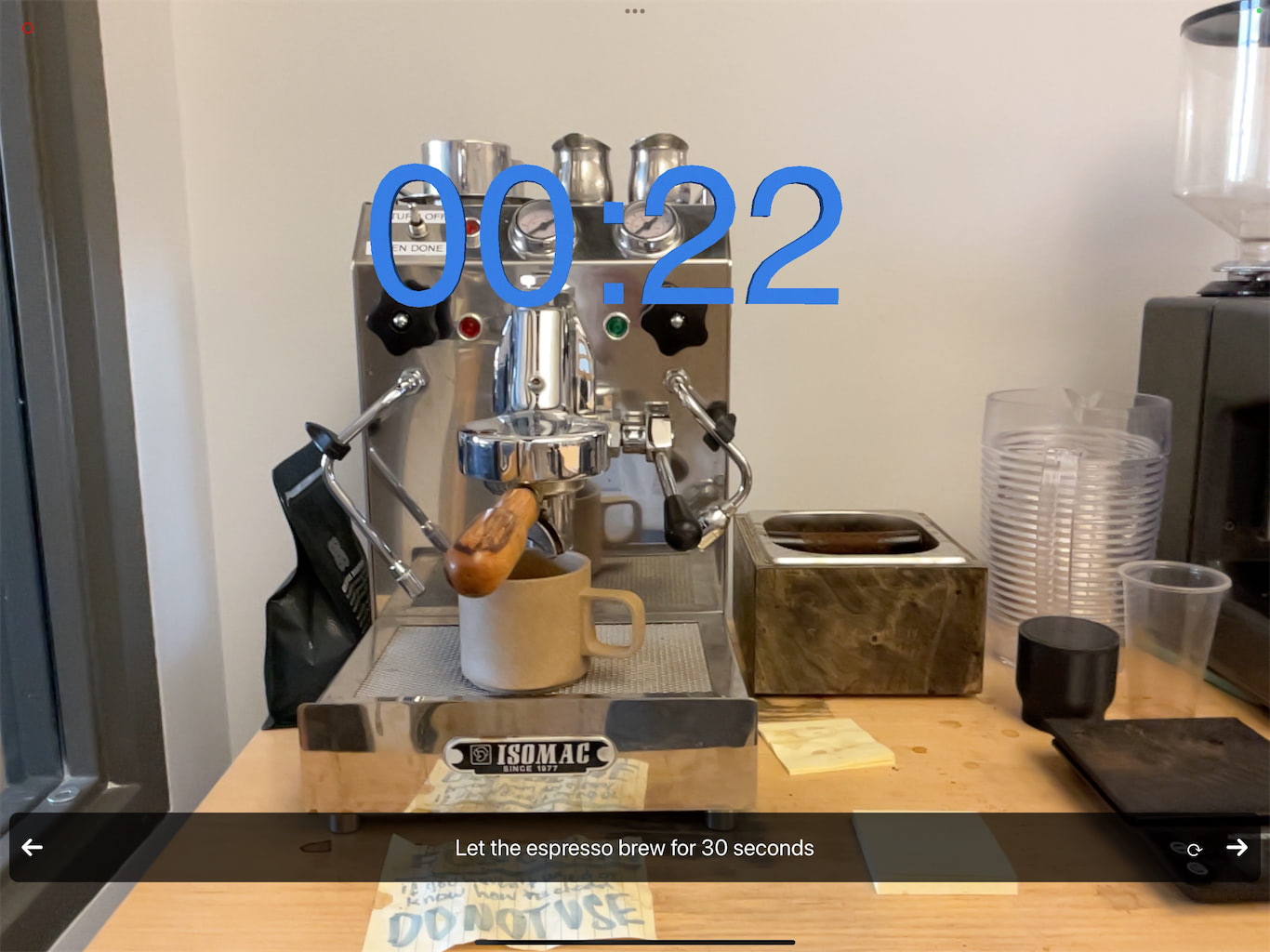}
\includegraphics[width=0.19\linewidth]{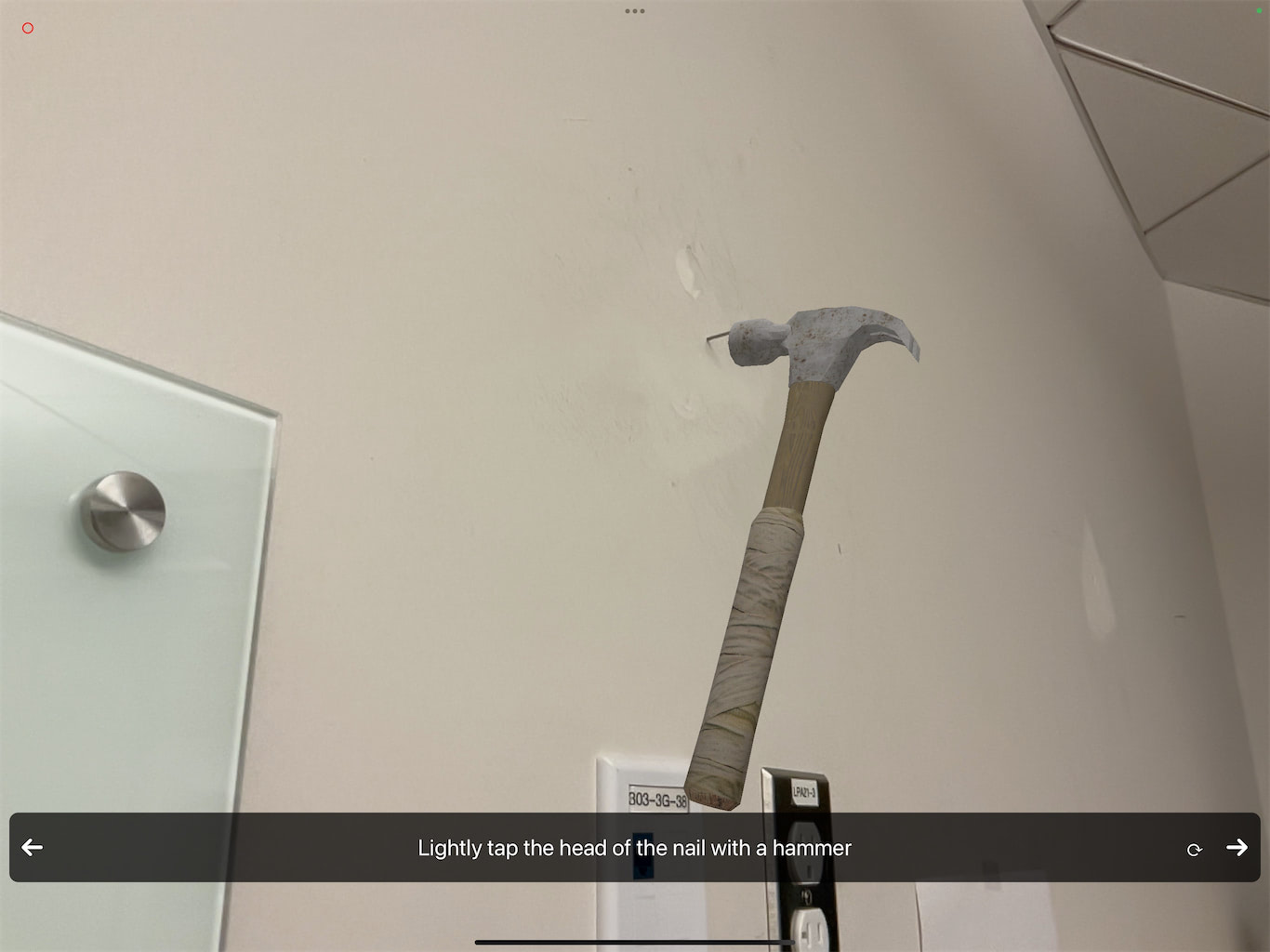}
\includegraphics[width=0.19\linewidth]{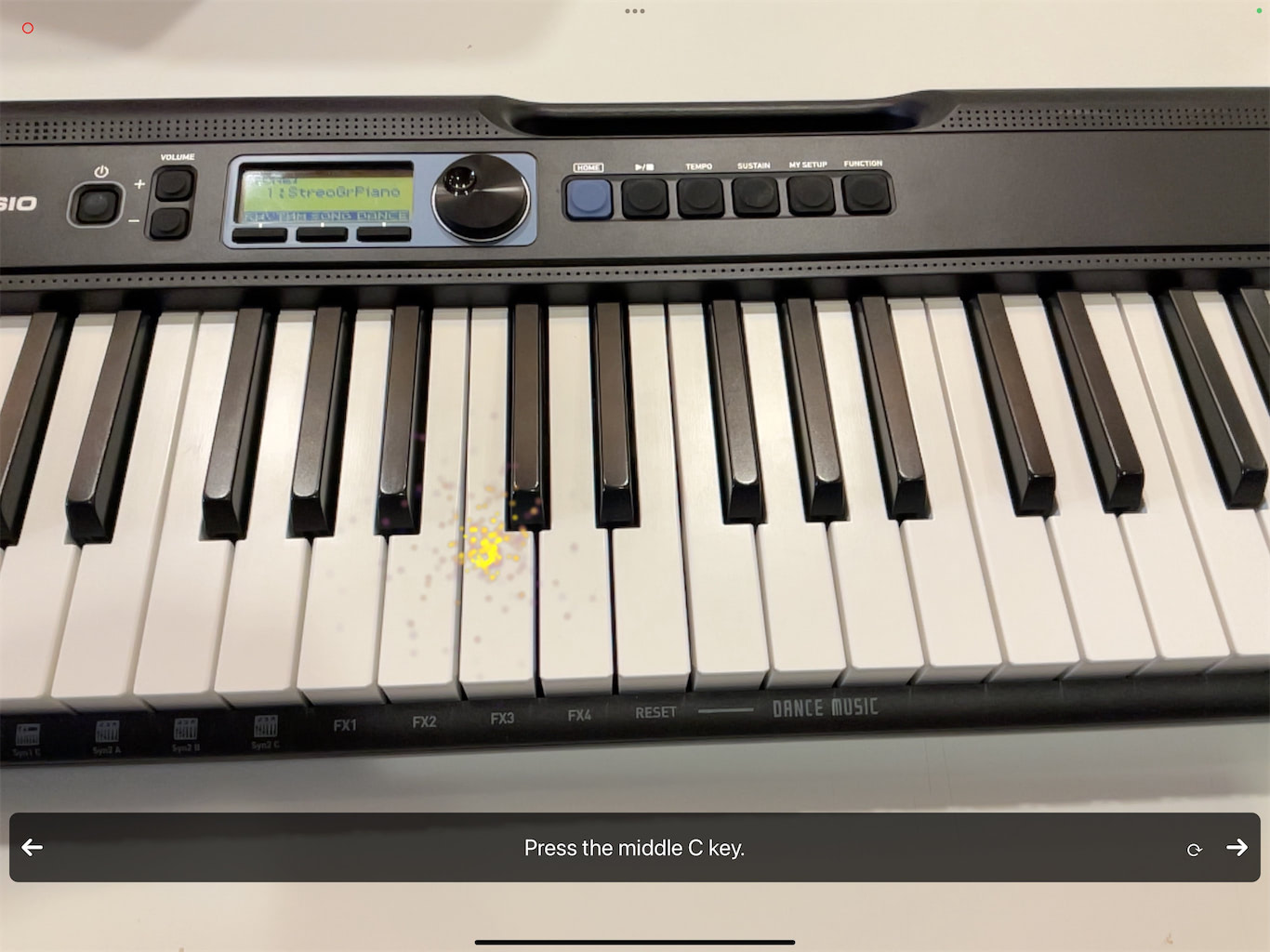}
\includegraphics[width=0.19\linewidth]{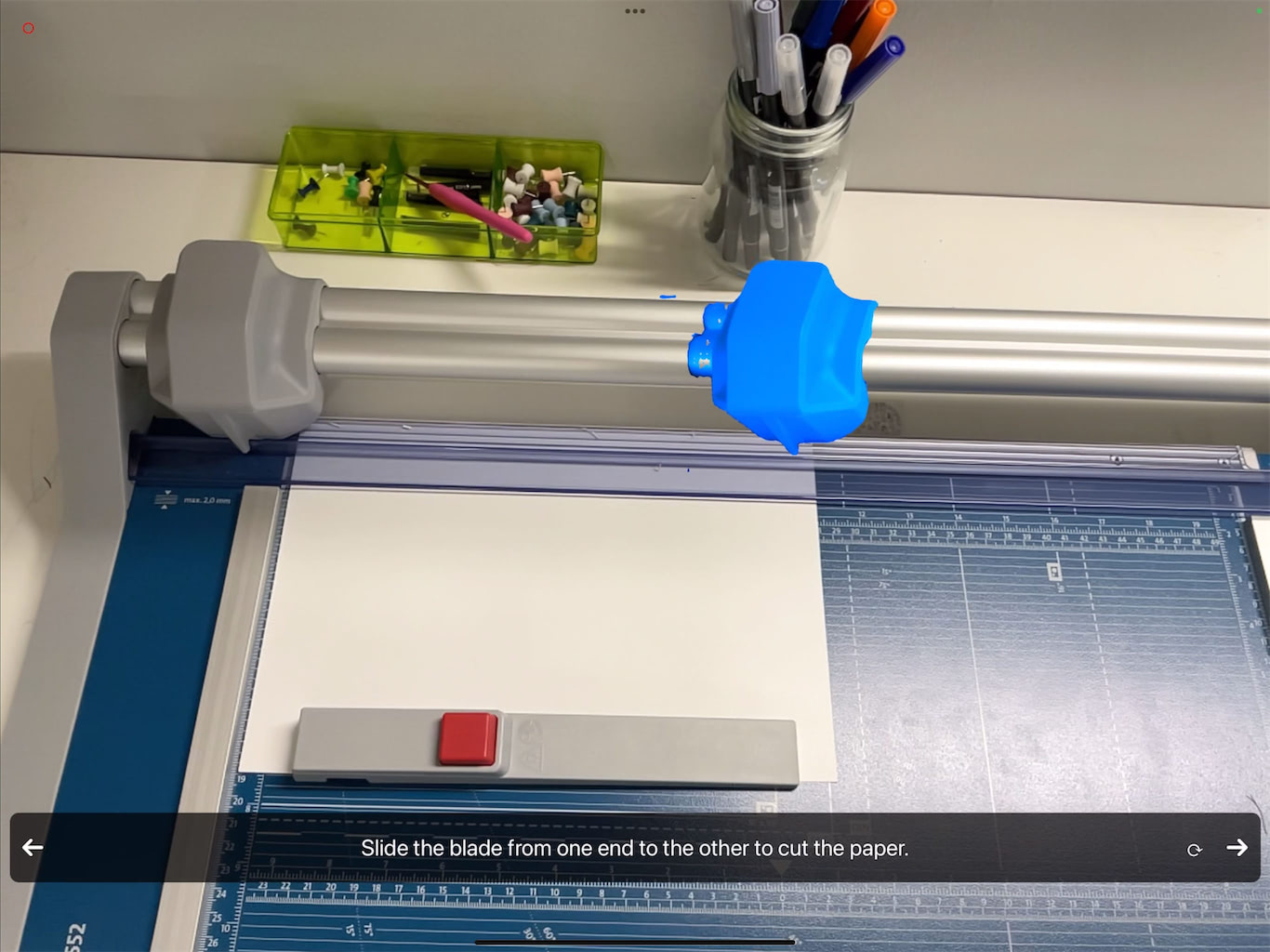}
\includegraphics[width=0.19\linewidth]{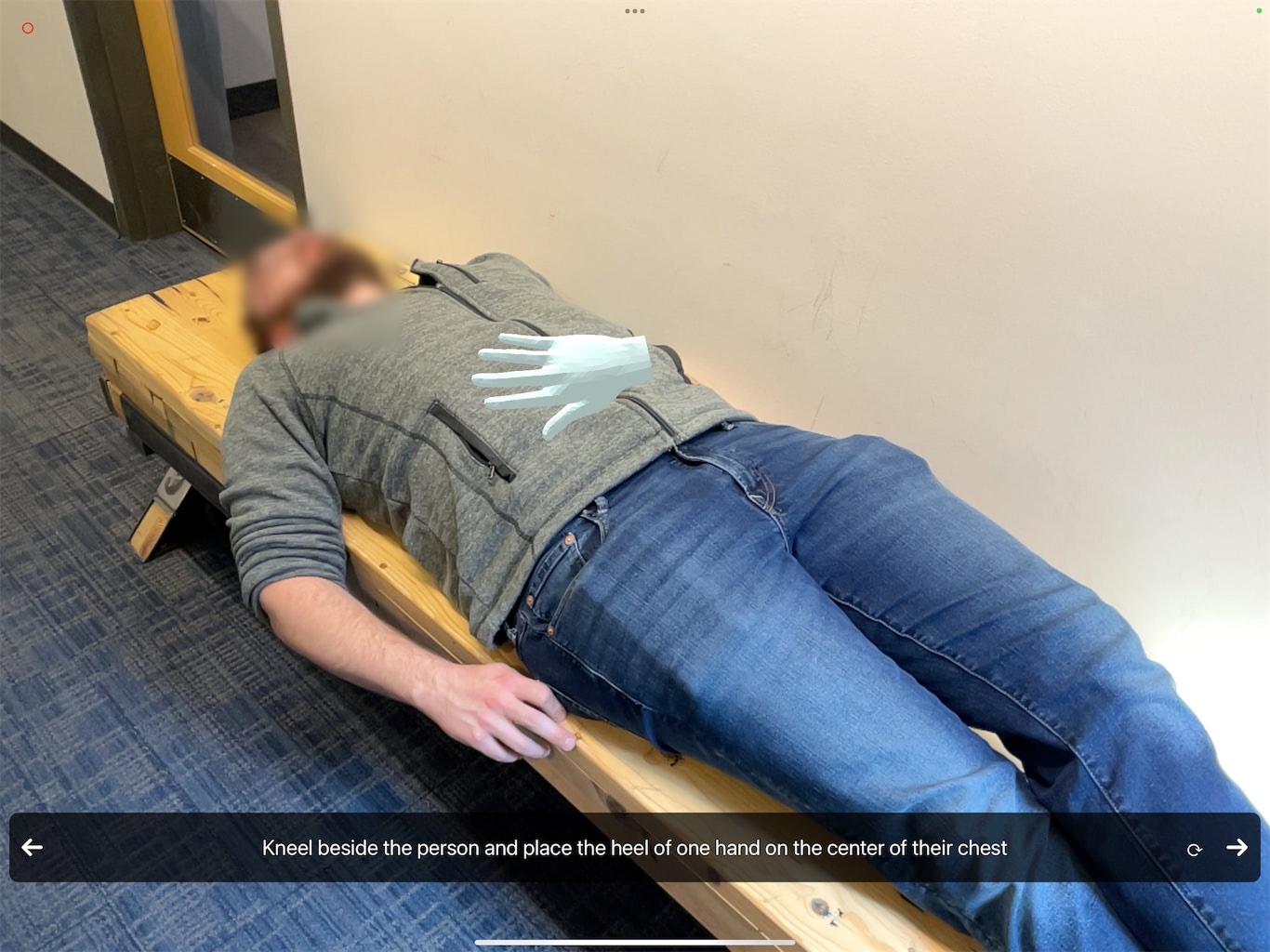}
\caption{Examples of diverse real-world tasks explored by users during the in-the-wild study.}
\label{fig:in-the-wild2}
\end{figure*}

\subsubsection{In-the-wild Exploration}
In the second phase of the study, participants were encouraged to freely explore various locations within our building, including the community kitchen, meeting rooms, research labs, and lounges. This phase was designed to assess the expressiveness and flexibility of the system in supporting user-driven tasks in real-world contexts. Participants were first guided through the process of entering a question to initiate an instruction session, after which they were free to choose and attempt tasks in the spaces around them. At the end of the session, we conducted a brief interview to gather their overall impressions of the system. This phase lasted approximately 15 minutes.

\subsection{Results}
\begin{figure}[h]
\centering
\includegraphics[width=0.98\linewidth]{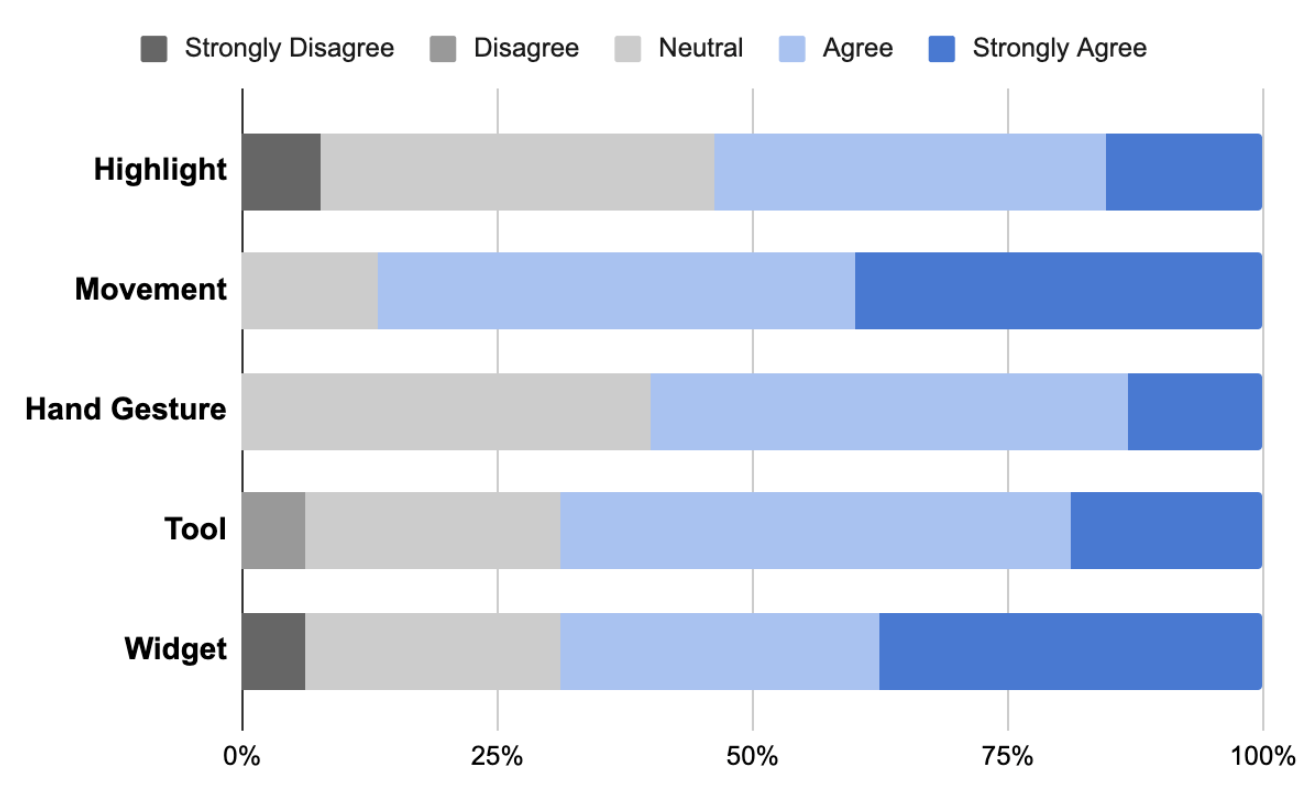}
\caption{Participant feedback on effectiveness of five visual guidance types on a 5-point Likert scale}
\Description{A diagram showing the placement of arrows for rotational movement guidance.}
\label{fig:evaluation}
\end{figure}

\subsubsection{System Usability and Visual Consistency}

We evaluated system usability using the System Usability Scale (SUS). Our system received a mean SUS score of 84.38 (SD = 12.17), indicating strong usability. The baseline condition received a score of 81.41 (SD = 7.23). While the overall SUS scores were not significantly different, our system scored higher on Q4 (need for technician: 89.62 vs. 73.44, p = 0.08) and Q9 (confidence: 81.25 vs. 67.19, p = 0.06). On the other hand, our system scored slightly lower on Q6 (consistency: 78.13 vs. 82.81, p = 0.11). This aligns with the qualitative feedback we received.

Participants occasionally raised concerns about inconsistency. P2 noted, “\textit{I don't know what visual I should expect. Sometimes it doesn't match my expectation and it took me a while to notice.}” Similarly, P15 said, “\textit{I like the blue highlight and I want to see it in all steps.}” Two participants (P7 and P10) missed the red bounding box on the 3D printer knob due to low contrast between the visual and the background (Figure~\ref{fig:3d-printer}.5). P11 suggested, “\textit{I like the dynamic particle effects more than the red box because it's easier to notice. Maybe the bounding box can also blink.}”

These findings highlight the importance of visual consistency and salience in building trust and ease of use. While participants were generally positive about the system's performance, they emphasized the need for more consistent and noticeable visual design across different steps.

\subsubsection{Completion Time and Error Rate}

Participants using our system made fewer errors overall, with error rates decreasing from 17.14\% to 10.42\% in the cleaning printer scanning area task and from 9.8\% to 7.7\% in the resetting 3D printer task. Completion times were comparable across conditions, with users of our system taking slightly longer (2:42 vs. 2:32 and 3:33 vs. 3:32), suggesting a modest tradeoff between speed and improved task accuracy.

Step-level analysis revealed that participants using our system often spent more time observing visual guidance animations—such as the wiping animation shown in Figure~\ref{fig:teaser}.3. At the same time, they made fewer errors compared to the baseline condition. For example, in the lifting ADF step in Figure~\ref{fig:teaser}.2, baseline participants sometimes touched the paper tray, whereas users of our system were more successful in interpreting the need to lift the entire component and correctly approach the handle. The next section will cover more information about the visual guidance effectiveness.

\subsubsection{Effectiveness of Visual Guidance Types}

Participants rated the effectiveness of each visual guidance type on a 5-point Likert scale. Figure~\ref{fig:evaluation} summarizes the average ratings.

Movement (M = 4.27, SD = 0.70) guidance was the most favored, particularly for illustrating large object manipulations. Participants noted that it conveyed not just the location of an interaction, but also the direction and manner of movement—capabilities that the baseline arrow lacked. However, Movement visuals were less effective when the system occasionally misidentified the target position in moving the 3D printer bed.

Widgets (M = 3.94, SD = 1.12) were praised for being clear and interactive. Participants liked their responsiveness and suggested they could be even enhanced by auto-advancing to the next step when an action is completed, or by linking with smart devices. Compared to the arrow, P3 mentioned "\textit{I really appreciate the timer especially when it is more than 10 seconds in the 3D printer task.}"

Tool (M = 3.81, SD = 0.83) guidance was appreciated for helping participants identify the correct tool and understand how to use it. The 3D model helped most participants identify the correct tool in their environment. However, if the model did not look exactly the same, novice users still needed assistance to confirm they had selected the right tool. They also requested more precise placement cues. P13 mentioned “I would like to see the tip of the scraper near the edge of the printed object to demonstrate the action.” To improve alignment with real-world tasks, P1 and P11 also suggested combining the tool animation with the blue segmentation to mark interaction zones.

Hand Gesture (M = 3.73, SD = 0.70) cues were particularly helpful for users with little prior experience, as they visually demonstrated physical hand actions. However, some participants took time to recognize the intended meaning of the gestures. In particular, the “hook” gesture in the printer task was occasionally misinterpreted as a pointing gesture, leading participants to focus on the interaction area rather than the motion itself, which caused delays in understanding how to perform the action. For example, P16 spent time pressing and sliding the lever before realizing the hook motion was required to unlock it. He reflected “I didn't realize the hand represents the gesture I should use, but once I noticed it, I should be able to follow the guidance easily.”

Highlight (M = 3.54, SD = 1.05) was the least effective among five visual types. While several participants felt the static bounding box was comparable to the baseline yellow arrow, many responded favorably to dynamic particle effects. P10 and P15 specifically noted that animated particles not only helped them more easily locate the target area but also evoked a sense of familiarity and enjoyment, reminding them of visual cues in video games.

\subsubsection{Impressed by In-the-wild Performance}
Participants were generally impressed by the system’s performance in real-world scenarios. Figure~\ref{fig:in-the-wild2} shows several examples from our in-the-wild user study, where users applied the system across a wide range of tasks: safety and emergency procedures, household appliances, creative and artistic tasks, software interfaces, and musical instruments. The system’s ability to generate relevant, context-aware guidance across such diverse domains demonstrated its versatility and robustness.

\section{Instructor Interview}
\begin{figure*}
\centering
\includegraphics[width=0.24\linewidth]{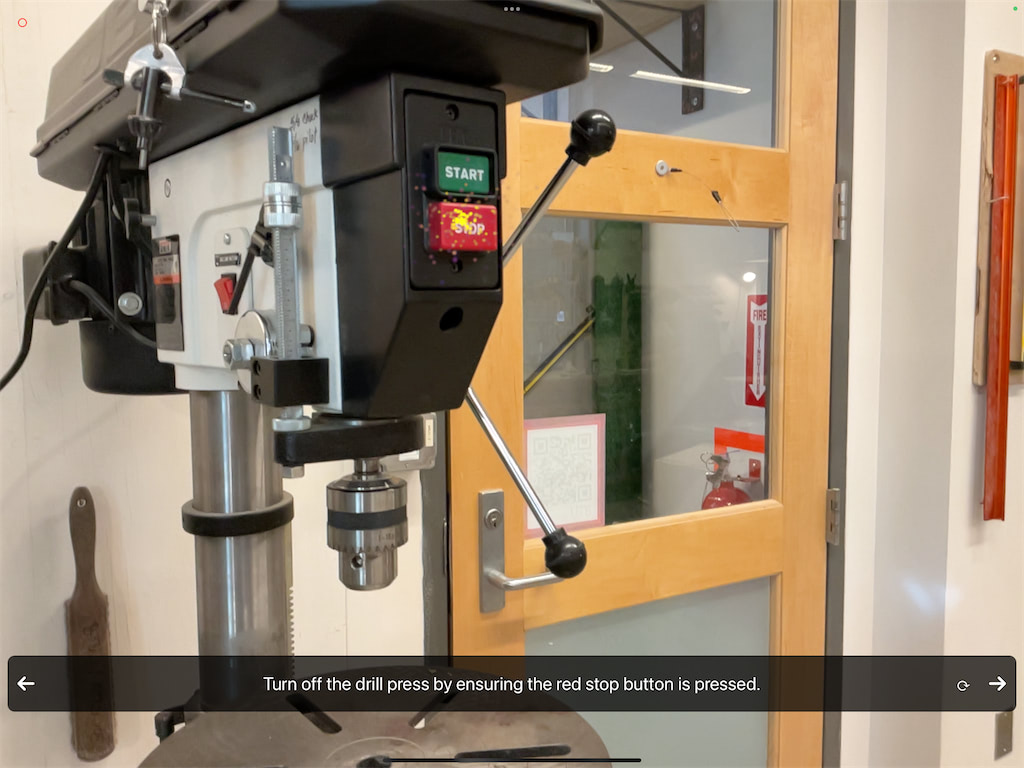}
\includegraphics[width=0.24\linewidth]{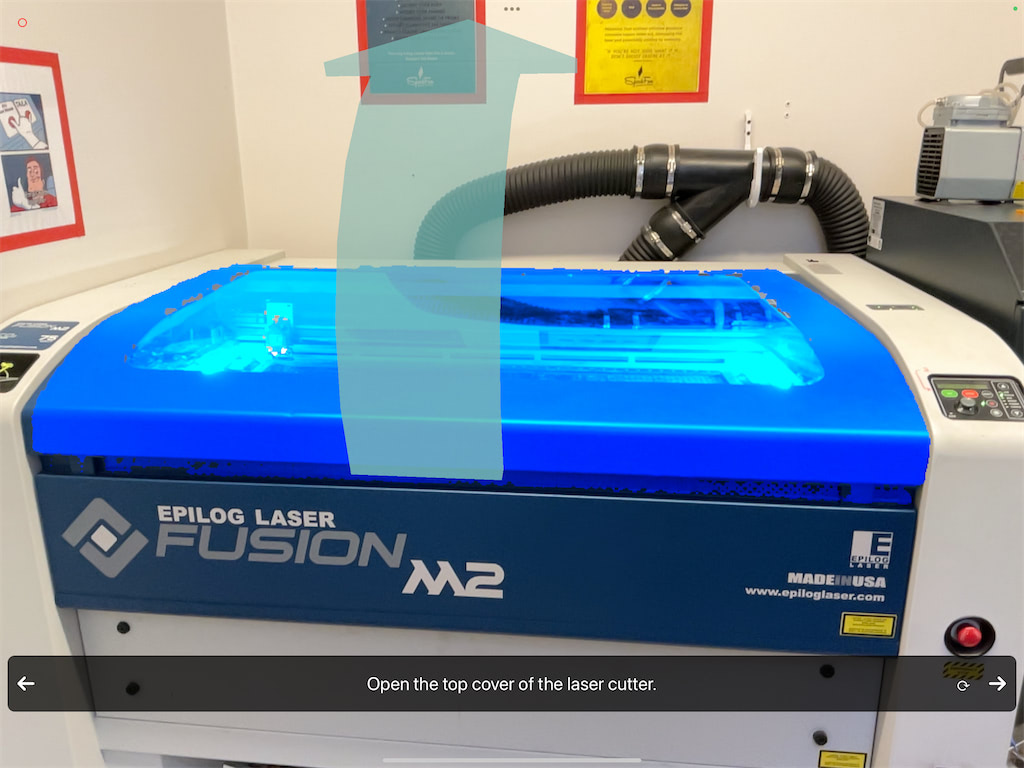}
\includegraphics[width=0.24\linewidth]{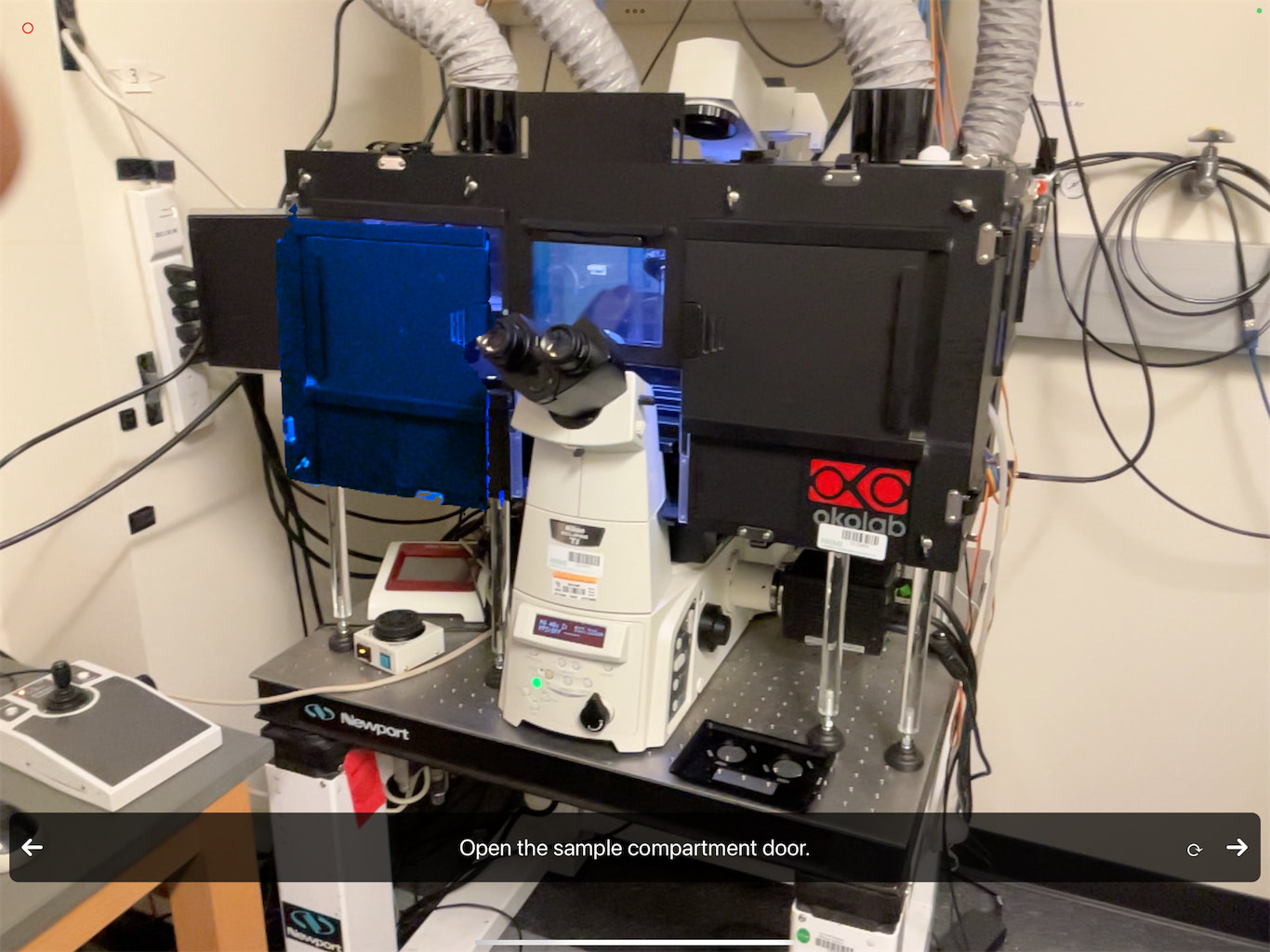}
\includegraphics[width=0.24\linewidth]{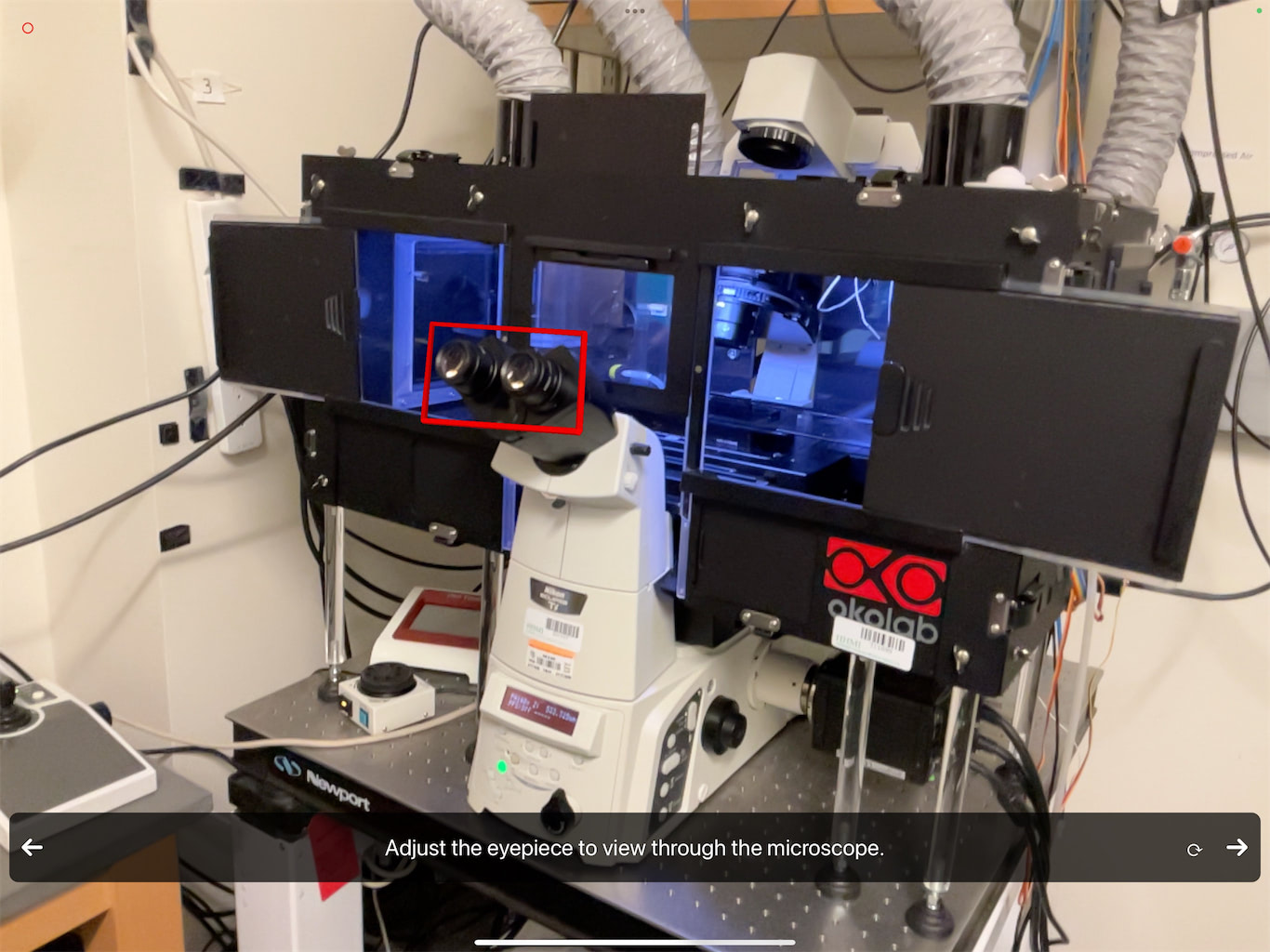}
\\[0.05cm]
\includegraphics[width=0.24\linewidth]{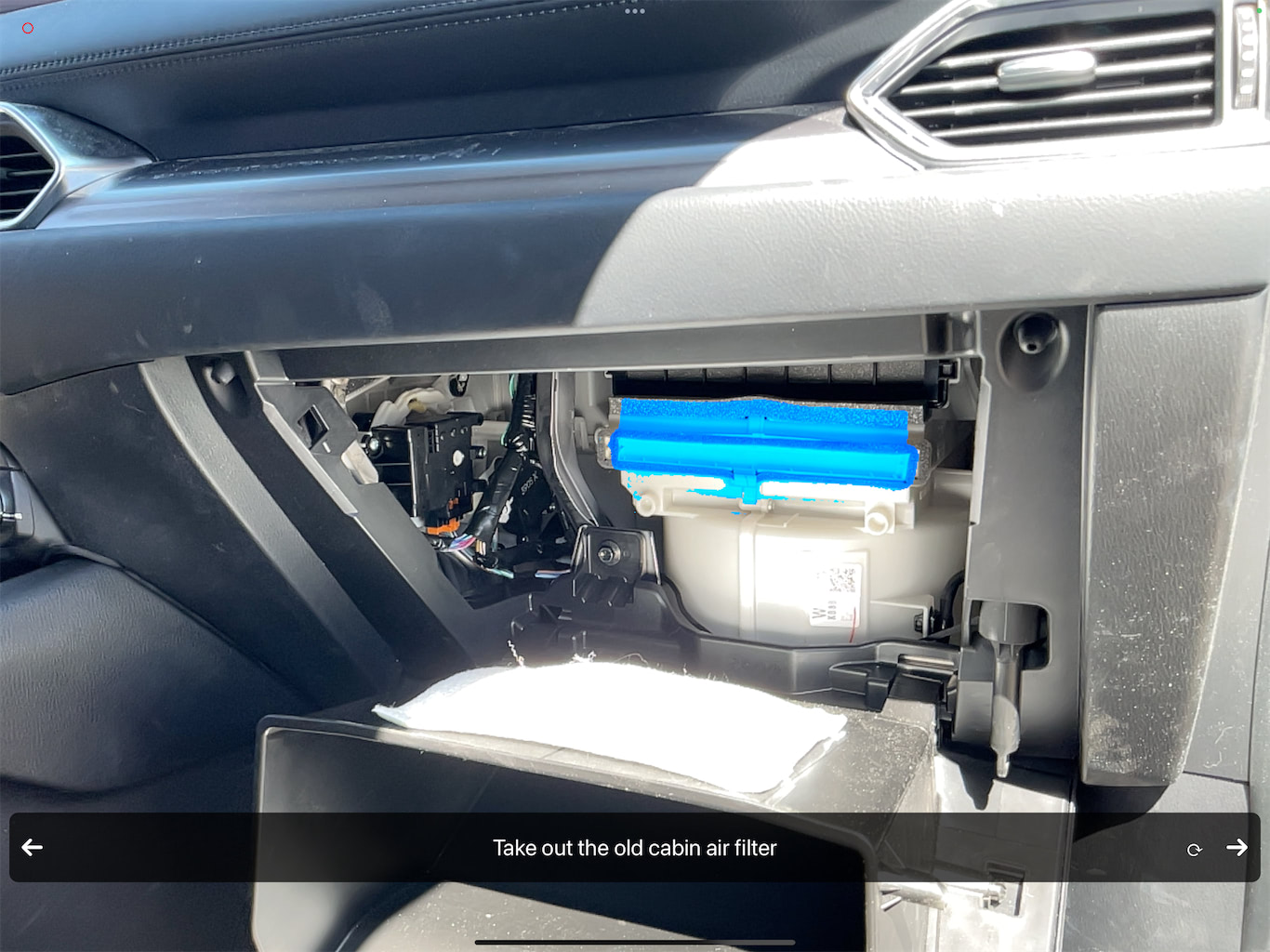}
\includegraphics[width=0.24\linewidth]{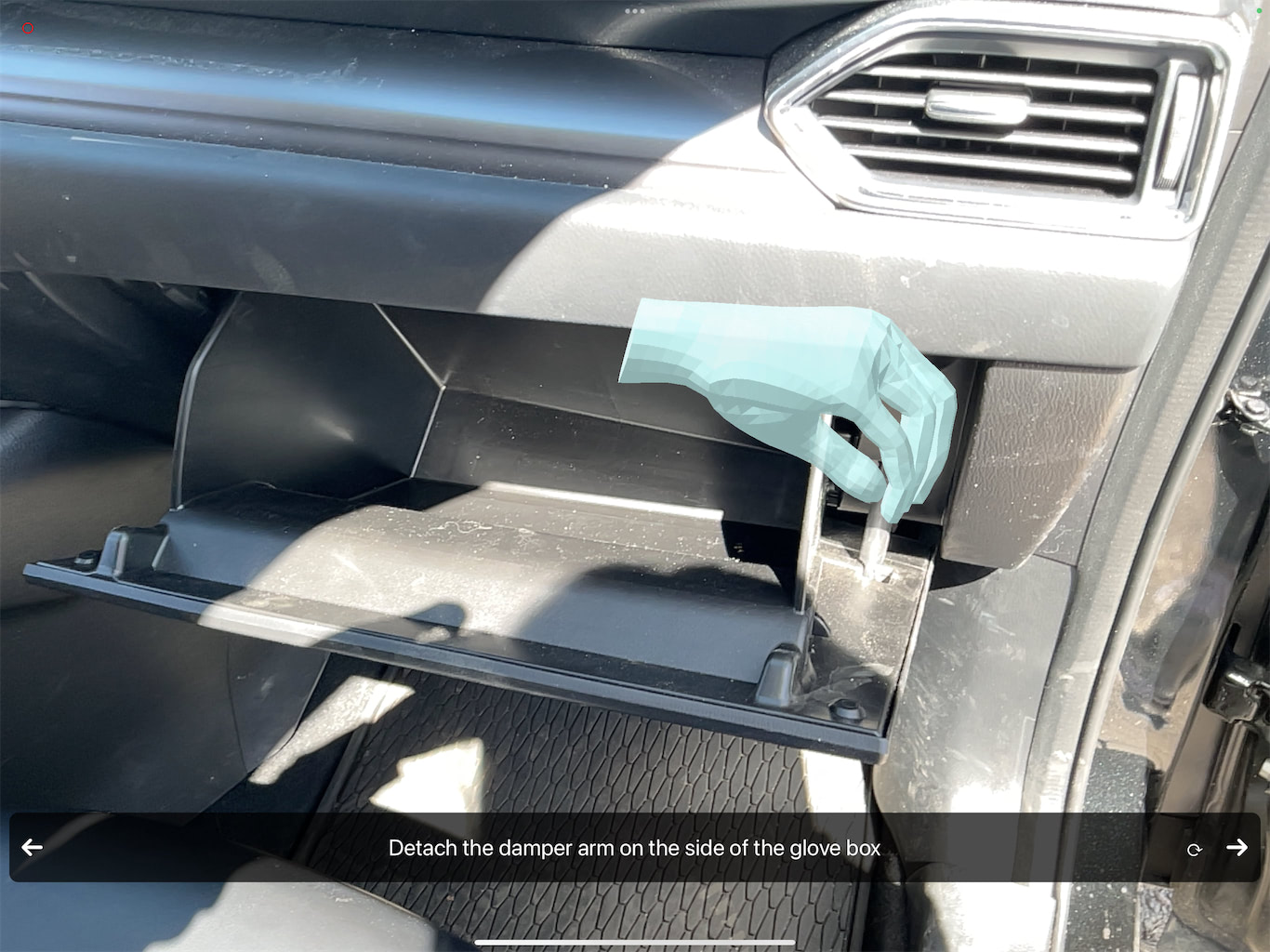}
\includegraphics[width=0.24\linewidth]{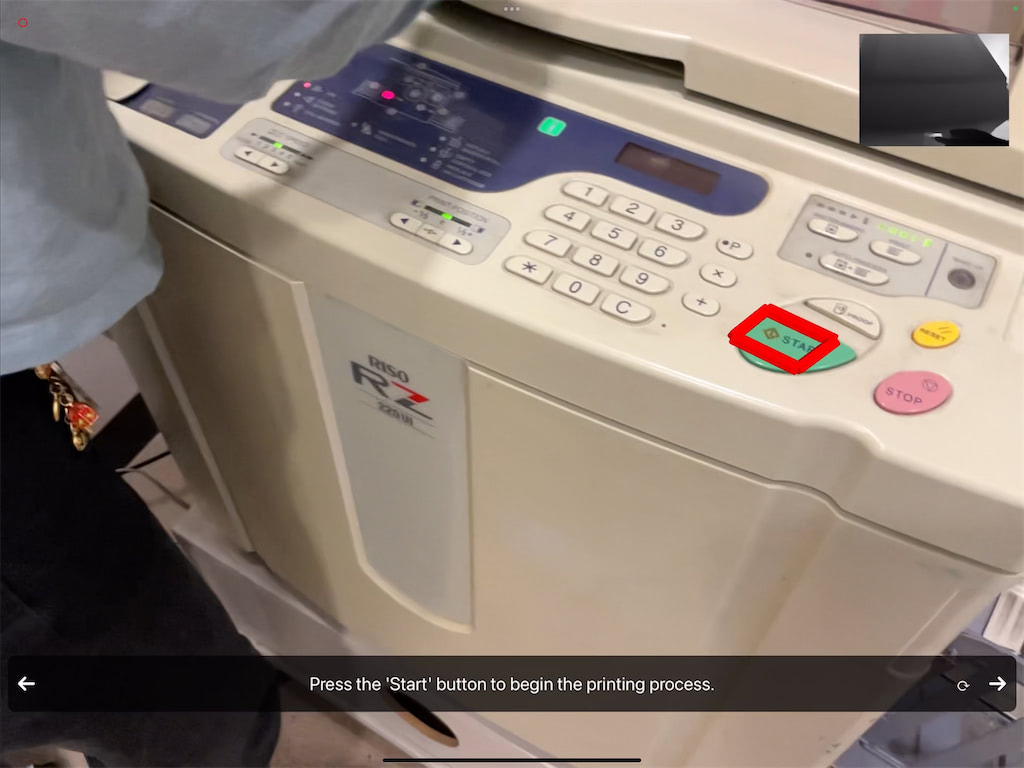}
\includegraphics[width=0.24\linewidth]{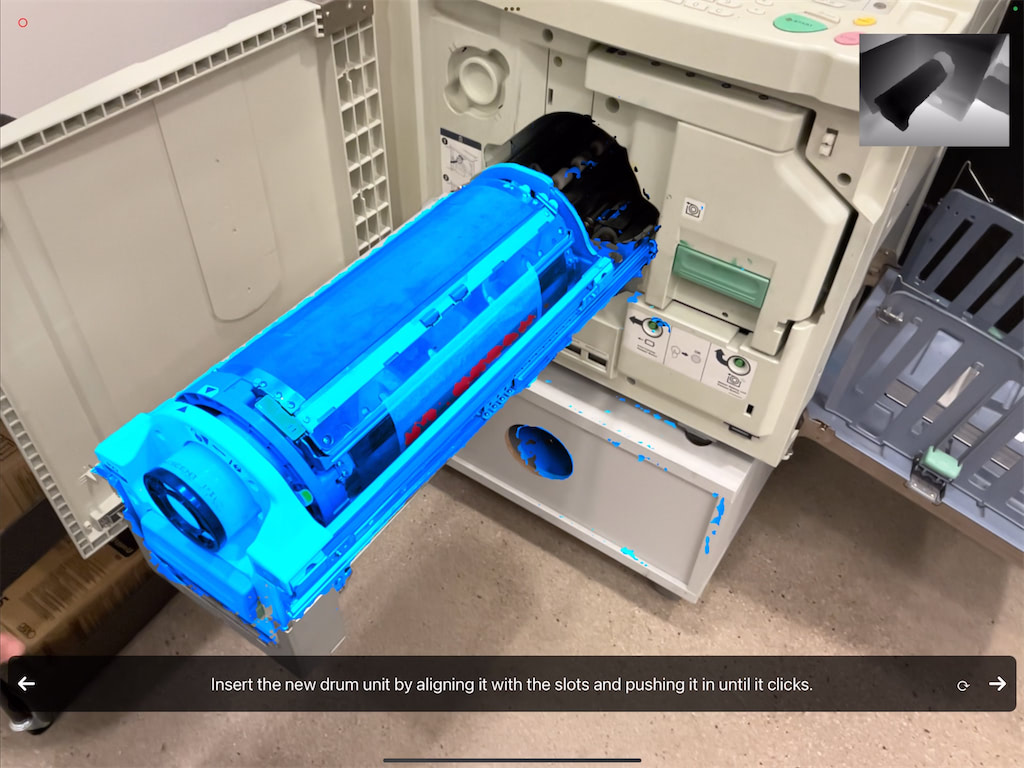}
\caption{Examples of tasks demonstrated during expert interviews}
\label{fig:expert}
\end{figure*}
In addition to evaluating \system with end users performing everyday physical tasks, we conducted semi-structured interviews with instructors who regularly provide hands-on training with machinery. The objectives of these interviews were to collect feedback from an instructor's perspective, explore how the system could be integrated into existing teaching workflows and identify areas for enhancement. We recruited four instructors (I1–I4) through personal connections, whose teaching topics ranged from maker space tools \& machines, microscopy, car maintenance and the use of RISO duplicators for artistic production (Figure \ref{fig:expert}). Their teaching experience spans from 1 to 20 years, with an average of 8 years. Depending on the equipment and learner needs, their training formats include both group workshops and one-on-one sessions.

\renewcommand{\arraystretch}{1.4}
\begin{table}[h]
\centering
\begin{tabular}{@{}l l l l@{}}
\toprule
\textbf{Instructor} & \textbf{Subject} & \textbf{Format} & \textbf{Experience} \\
\midrule
I1 & Maker Space       & Group Workshop & 20 \\
I2 & Microscope         & One-on-One     & 9  \\
I3 & Car Maintenance    & One-on-One     & 1  \\
I4 & RISO Duplicator    & Group Workshop & 2  \\
\bottomrule
\end{tabular}
\vspace{0.5em} % adjust value as needed
\caption{Overview of instructor backgrounds, including training topic, teaching format, and years of experience.}
\label{tab:instructor-background}
\end{table}

\subsection{Methods}
The interviews were conducted in person at the instructors' regular training spaces. We started by asking about their current instructional practices, including the training subject, teaching format, duration of the session, and the types of questions that learners typically ask before, during and after the session. This introductory phase lasted approximately 15 minutes.

Next, we introduced \system with a live demonstration and invited the instructors to simulate a training session by guiding one of the researchers. During this hands-on phase, instructors were encouraged to input relevant questions into the system as they arose from either the researcher or previous experience, and observe the visual guidance generated in response. They were free to explore and interact with as many examples as they wanted. This exploration phase was limited to 20 minutes.

Finally, we conducted a semi-structured interview lasting 15 minutes to reflect on their experience with the system. Through open-ended questions, we gathered feedback on usability, perceived value, and how \system could be integrated into their existing teaching practices.

\subsection{Results}
\subsubsection{Personalized Learning}
Instructors I1 and I4 both conducted their training sessions as group workshops. They noted that the system could provide students with a more personalized learning experience tailored to their purposes after the group session. For instance, I4 shared, "\textit{Some students want to explore a mix of colors, so they need to change the color drum. We don't usually cover this in the group workshop, so they may use this tool to learn.}"

\subsubsection{Troubleshooting}
I4 highlighted the value of the system for troubleshooting, stating, "\textit{In the past, I had to jump between the part diagram page and the error code page to understand how to fix it. This tool highlights the components on the actual machine. I see this as a valuable tool for my students and myself when troubleshooting error codes.}" I1 added that students often struggle with debugging circuits and commented, "\textit{This tool would be great if it can tell students this wire is connected wrong.}"

\subsubsection{Reminding Students of Standard Procedures}
While I4 noted that the system might be less applicable for artistic tasks—where understanding principles and experimentation are key—both I1 and I2 viewed it as a useful tool for reinforcing standard procedures. I2 shared that they had printed a paper with five steps for turning on a microscope, yet students still occasionally missed a step. "\textit{Maybe they can follow along with this system. The system is like a reminder,}" I2 explained. However, since the system is generative, it does not guarantee consistency. Both instructors expressed interest in being able to manually input text instructions, seeing the system as a potential authoring tool.

\subsubsection{Bi-directional Interaction to Refine the Guidance}
Instructors emphasized the need for bi-directional interaction, where both the user and the system can ask questions to refine and adapt task guidance in real time. I3 highlighted that learners may encounter ambiguous steps mid-task and need to ask for clarification—for example, a step like “\textit{remove the cabin from its hinge}” actually involves detaching a damper arm and squeezing both sides of the cabin, which is not obvious from the initial instruction. In the other direction, I2 pointed out that the system sometimes requires additional input from the user to generate accurate guidance. For instance, when a student asks, “\textit{How do I load the sample into this microscope?}”, the system must first determine which immersion technique the user wants to use, as this choice significantly alters the procedure.

%% file: 7-future-work.tex
\section{Limitations and Future Work}
\subsection{Integration with Head-Mounted Displays}

While the current implementation of Guided Reality runs on a mobile AR interface, we recognize that head-mounted displays offer advantages, especially for two-handed tasks. The iPad provides mobility and accessibility, but we observed that users often shift attention between the screen and the physical workspace, which can disrupt task flow. We plan to port the system to the Meta Quest in the near future. Rather than replacing mobile AR, our goal is to expand platform support and leverage the strengths of both devices across diverse use cases.

\subsection{Generative Models for Functional 3D Guidance}

To move toward fully generative guidance, we explored 3D model generation using tools like Meshy~\cite{MeshyAI_Text_to_3D}. While these models can serve visual purposes, they often lack a grounded understanding of functional orientation and interaction context. For instance, current generative models may produce plausible geometries but fail to orient the model as expected even when prompted explicitly. Future work could investigate hybrid approaches that combine generative geometry with explicit spatial constraints or rotate the model afterwards with an understanding of its initial pose. Similarly, recent advances such as Text2HOI~\cite{cha2024text2hoi}, which generate 3D hand-object interactions from textual descriptions and mesh data, present a promising direction for expanding gesture diversity and realism in future iterations of the system. However, we expect that further refinement will be necessary to ensure the generated outputs are not only visually convincing but also functionally accurate and contextually grounded as with prior work such as CARING-AI~\cite{shi2025caring}.

\subsection{Guidance for Out-of-View Components}

The current system assumes that all relevant components are visible within the camera view when visual guidance generation pipeline is triggered. However, in real-world settings, key components (e.g., a power button located on the back of a machine) may lie outside the user’s field of view. As future work, we plan to implement continuous sampling of the camera feed to track component positions and anchor them persistently in space. Additionally, we are interested in enabling the system to output and interpret relative spatial terms to guide users to reposition themselves and locate hidden components, such as the power button on the back of the machine.

%% file: 8-conclusion.tex
\section{Conclusion}
We presented Guided Reality, a fully automated system that generates visually enriched, spatially situated AR guidance by integrating LLMs and vision models. Our approach advances prior AR task guidance systems by introducing a novel strategy for selecting and embedding five distinct types of visual augmentations, each grounded in a systematic analysis of user manuals. Our user study revealed that users found the system usable and effective, especially praising the movement and dynamic widget types for their clarity and relevance. In-the-wild testing further highlighted the system’s robustness and adaptability across real-world contexts. Interviews with instructors emphasized its potential not only as a just-in-time training tool but also as a platform for supporting personalized learning, troubleshooting, and reinforcing standard procedures.

%% file: acknowledgements.tex
\begin{acks}
This research was partially funded by the JST PRESTO Grant Number JPMJPR23I5.
\end{acks}